\documentclass[aps,twocolumn,superscriptaddress,prb,showpacs]{revtex4-1}
\usepackage{graphicx}
\usepackage{amsmath}
\usepackage{epstopdf}

\newcommand{\Wigner}{Wigner Research Center for Physics and Optics, Hungarian Academy of Sciences, P.O. Box 49, H-1525 Budapest, Hungary}
\newcommand{\Imomec}{IMOMEC division, IMEC, Wetenschapspark 1, B-3590 Diepenbeek, Belgium}
\newcommand{\Imo}{Institute for Materials Research (IMO), Hasselt University, Wetenschapspark 1, B-3590 Diepenbeek, Belgium}
\newcommand{\BPUni}{Department of Atomic Physics, Budapest University of Technology and Economics, Budafoki\'{u}t 8., H-1111 Budapest, Hungary}

\begin{document}
\title{
Identification of nickel-vacancy defect in the photocurrent spectrum of diamond by means of \emph{ab initio} calculations
}
\author{E. Londero}\affiliation{\Wigner}
\author{E. Bourgeois}\affiliation{\Imomec}\affiliation{\Imo}
\author{M. Nesladek}\affiliation{\Imomec}\affiliation{\Imo}
\author{A. Gali}\affiliation{\Wigner}\affiliation{\BPUni}

\date{\today}

\begin{abstract}
There is a continuous search for solid-state spin qubits operating at room temperature with excitation in the IR communication bandwidth. Recently we have introduced the photoelectric detection of magnetic resonance (PDMR) to read the electron spin state of nitrogen-vacancy (NV) center in diamond, a technique which is promising for applications in quantum information technology. By measuring photoionization spectra on a diamond crystal we found two ionization thresholds that were not reported before. On the same sample we also observed absorption and photoluminescence signatures that were identified in literature as Ni associated defects. We performed \emph{ab initio} calculation of the photo-ionization cross-section of the nickel split vacancy complex (NiV) and N-related defects in their relevant charge states and fitted the concentration of these defects to the measured photocurrent spectrum, which led to a surprising match between experimental and calculated spectra. This study enabled to identify the two unknown ionization thresholds with the two acceptor levels of NiV. Because the excitation of NiV is in infrared, the photocurrent detected from the paramagnetic NiV color centers is a promising way towards designing a novel type of electrically readout qubits.  
\end{abstract}


\maketitle


Diamond is rich in point defects that are able to change the optical and electrical properties of the host material, often called, color centers~\cite{Zai01}. One of the prominent color centers is the nitrogen-vacancy (NV) defect in diamond. In particular, the negatively charged NV (NV$^-$) color center acts as a solid state quantum bit~\cite{jel06, childress06} which can be employed in diverse nanoscale sensing~\cite{maze08, balasubramanian08, rondin14, dolde14, neumann13, toyli13} and quantum communication applications~\cite{maurer12, pfaff14, hensen15}. These quantum applications of NV center are based on the long spin coherence time of NV center~\cite{bala09, jel06} and the optical readout and initialization of the spin state~\cite{gruber1997}. Recently, the spin state of NV($-$) has been detected by photoelectric detection of magnetic resonance (PDMR)~\cite{bourgeois2015}. In this scheme, the electron spin state is initialized optically, then illumination is applied to promote the electron to the conduction band, and finally the resulting 
photocurrent is detected. It has been found that the photoionization is spin-selective, that is the base of PDMR readout of the spin~\cite{bourgeois2015}. Recent advances on the enhancement of the PDMR contrast for readout~\cite{brandt2016, bourgeois2016, gulka2017} pave the way toward the detection of single NV centers. The detected electron rate of PDMR is significantly higher than the photon rate of optically detected magnetic resonance~\cite{bourgeois2015}. In recent experiments, by applying dual beam photoexcitation~\cite{bourgeois2016}, we were able to increase further the photoelectric detection rate of NVs. In our efforts of studying the mechanism of photoionization of the NV defect~\cite{bourgeois2016} and measuring the photocurrent response of the diamond sample in a wide spectral range, two photocurrent bands with unknown origin appeared in the near-infrared (NIR) range at about 1.2~eV and 1.9~eV. The identification of these ionization bands is of high importance, at least for two reasons: (i) photoresponsive defects give unwanted background in the PDMR spectrum of NV$^-$ reducing the PDMR contrast, and therefore their identification guides the diamond materials fabrication to avoid their presence; (ii) these defects, that can be certainly photoionized, might turn to be advantageous for electrical detection of spin states as an alternative to the NV center if they have favorable magneto-optical properties. Specifically for electrical readout of spin center strong optical radiative transitions are not \emph{a priori} necessary, and novel, yet unidentified defects may outperform the NV quantum bits in diamond. Qubits enabling initialization in the near IR part of the spectrum are specifically interesting for quantum information applications.    

In this paper we apply density functional theory (DFT) to identify the origin of the photocurrent bands detected on an irradiated and annealed nitrogen-rich diamond, by comparing the calculated and experimental photocurrent spectra. We find that the bands originate from the two acceptor levels of the nickel split-vacancy complex, that we label NiV in this paper. We propose that the NiV complex can potentially act as a solid state quantum bit, and might be a subject for future PDMR studies in the NIR region for photoionization.


The experimental photocurrent spectra were taken from a type-Ib HPHT diamond plate that was irradiated by protons with energy of 6.5~MeV and dose of $1.13\times10^{16}$~cm$^{-2}$ and subsequently annealed for one hour at 900$^{\circ}$C under argon atmosphere. After this treatment, the sample was cleaned and oxidized. For photocurrent spectroscopy measurements, the sample was equipped with coplanar electrodes (inter-electrode distance: 100~$\mu$m). A monochromatic light (1 to 300~$\mu$W depending on the wavelength) produced using a halogen lamp associated to a monochromator was focused onto the sample. The free charge carriers induced by photoionization of defects in the diamond crystal were directed toward electrodes by applying a DC electric field of $5\times10^4$~V cm$^{-1}$. The resulting photocurrent was pre-amplified, measured by lock-in amplification and normalized to the flux of incoming photons.  
The total photocurrent spectrum contains the contributions resulting from the ionization of individual defects at a given concentration. The continuous wave photoexcitation leads to a steady state balance between the charge states of the defects that is different from the thermal equilibrium. As a consequence, the observed concentration of a defect in a given charge state may change when the conditions of illumination are varied. To better identify the origin of the defects responsible for photoionization bands, two methods were applied in the photocurrent spectroscopy experiments. In the first case (method 1), no blue bias light illumination was applied, whereas in the second case (method 2) a blue bias illumination (parameters: 2.4-3~eV excitation energy and 1.8~mW power) was used. The blue bias light allows to manipulate the occupation of deep defects, which is particularly useful for selectively promoting (or excluding) photoionization of specific defects. 

\begin{figure}[h]
\includegraphics[width=0.48\textwidth]{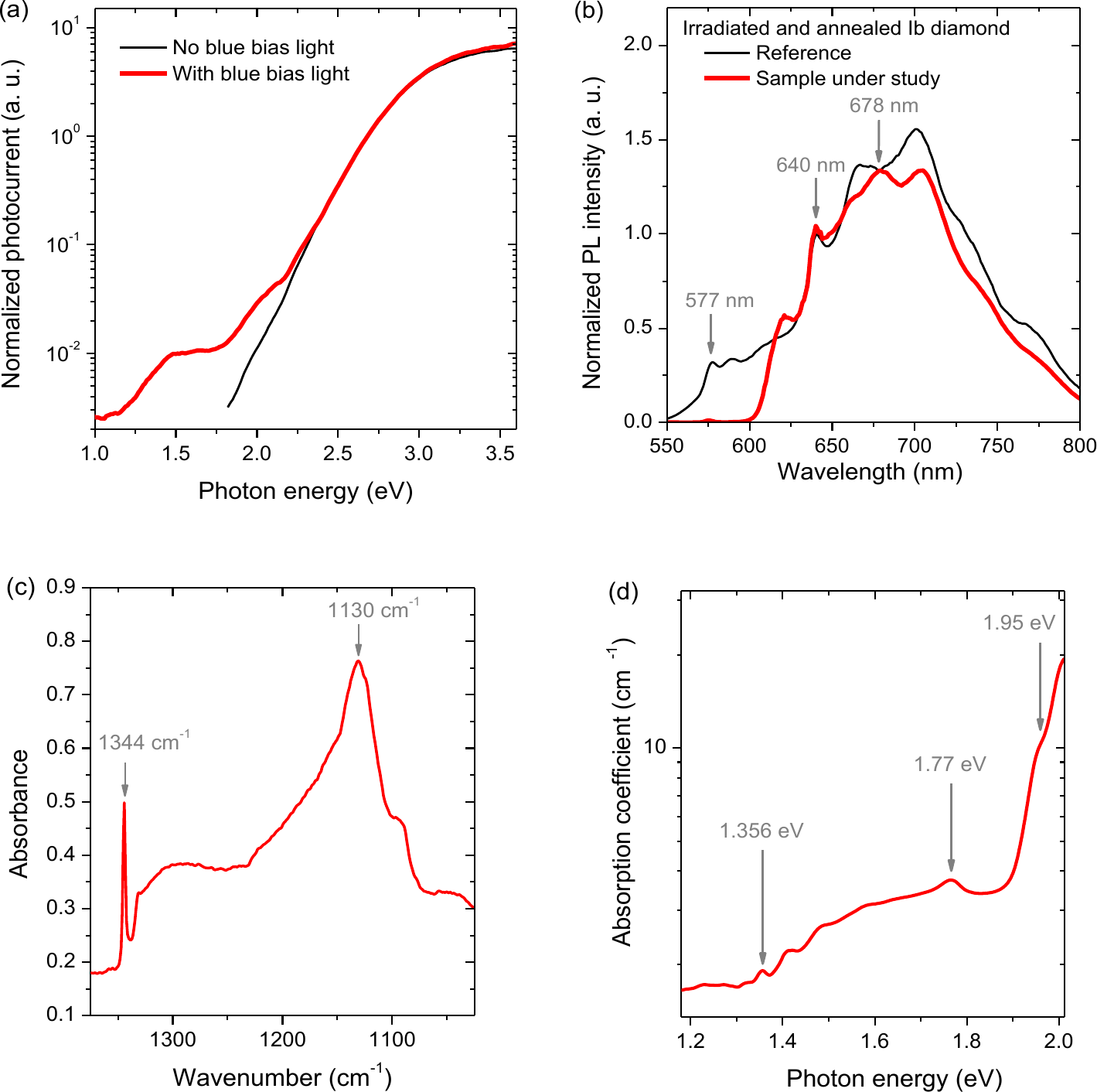}
\caption{\label{fig:exp}(a) Photocurrent spectra measured on irradiated and annealed type-Ib diamond without (method 1) and with (method 2) blue bias illumination. (b) Photoluminescence (PL) spectra (Excitation: 532~nm, 1.6~mW). The PL peak at 678~nm, observed in the spectrum of the sample under study (red curve) and attributed to a Ni-related defect, was not detected in the spectrum of a reference irradiated and annealed type-Ib diamond (shown for comparison, black curve). (c) FTIR spectrum. The absorption peaks of N$_\text{s}^0$ at 1344~cm$^{-1}$ and 1130~cm$^{-1}$ can be observed. (d) Optical absorption coefficient in the UV-visible range, measured at 77~K. The ZPL of 
NV$^-$ appears at 1.95~eV. The absorption peak detected at 1.77~eV, already reported on irradiated type-Ib diamond, has been attributed to an interstitial defect~\cite{Iakoubovskii01}. The ZPL at 1.356~eV and associated phonon replica at higher energies have been observed in diamond containing both N and Ni~\cite{Zai01}.}
\end{figure}

On the spectrum recorded without blue bias light (method 1) a broad ionization band with onset around 2.7~eV is observed [see Fig.~\ref{fig:exp}(a)]. Our previous work has shown that this feature resulted from a combination of the ionization bands of substitutional nitrogen (N$_\text{s}^0$), NV$^0$ and NV$^-$ (see Ref.~\onlinecite{bourgeois2016}). An additional band with onset around 1.9~eV can also be distinguished. In the presence of a blue bias light (method 2), the photocurrent associated with the 1.9~eV ionization band increases and a new photocurrent band with threshold around 1.2~eV emerges [see red curve in Fig.~\ref{fig:exp}(a)]. We note that red light illumination (1.75-1.95~eV excitation energy and 1.5~mW power) did not produce the same effects. 

To identify the defects potentially responsible for the unidentified photocurrent bands with onset at 1.2 and 1.9~eV, the sample was further characterized by photoluminescence (PL), Fourier transformed infrared absorption (FTIR), and optical absorption spectroscopy [see Fig.~\ref{fig:exp}(b)-(d)]. 
The concentration of NV$^-$ was determined to be  $[$NV$^-]$$\sim$34~ppm, 
by comparing the PL intensity measured in confocal mode to that measured using the same objective on a reference high-pressure high-temperature (HPHT) sample with calibrated NV$^-$ concentration. The concentration of NV$^0$ centers is evaluated by comparing the PL spectrum obtained under low continuous wave green light excitation to the spectra obtained on single NV$^-$ and NV$^0$ (see Ref.~\onlinecite{ron10}), which resulted in the $[$NV$^0]$$\sim$1.1~ppm. The concentration of N$_\text{s}^0$ was estimated from the absorption coefficient at 1130~cm$^{-1}$, determined by FTIR absorption spectroscopy~\cite{wee2007}, which yielded $[$N$_\text{s}^0]$$\sim$220$\pm$20~ppm. The absorption peak of N$_s^+$ at 1332~cm$^{-1}$ (Ref.~\onlinecite{lawson1998}) is not detected, indicating that in this sample the N$_\text{s}$ defect is in large majority in its neutral state.  These data are summarized as "dark-exp." in Table~\ref{tab:conc}. Beside the signatures of NV$^0$  -- photoluminescence observed below 639~nm -- and NV$^-$ defects -- zero-phonon-line (ZPL) at 639~nm and associated phonon side bands at higher wavelengths -- a broad peak is observed in the PL spectrum at a wavelength of 678~nm (1.83~eV). This peak is associated with a nickel defect~\cite{collins82, lowther1995, iakoubovskii2000, goss04, larico09, thiering}. Another evidence for the presence of nickel is the ZPL at 1.356~eV and associated phonon replica observed in the UV-visible absorption spectrum which were reported for synthetic diamonds containing Ni and N (see Refs.~\onlinecite{Zai01, Yelisseyev95}). The concentration of the Ni related defects cannot be determined form this data. 

The thresholds of the two unidentified ionization bands observed at 1.2~eV and 1.9~eV in photocurrent spectra closely resemble our calculation of the first and second acceptor levels of the NiV defect at $E_\text{V}$+1.22~eV and +1.90~eV~\cite{thiering}, respectively, where $E_\text{V}$ is the valence band maximum.
Although other Ni-related defects may exist in the sample, only the NiV$^0$ and NiV$^-$ show photoionization bands at $\approx$1.2~eV and 1.9~eV, respectively. 
Therefore, our working model is that the ionization of NiV$^0$ and NiV$^-$ are responsible for these features. To verify this hypothesis, we used DFT calculations to determine the photocurrent spectra obtained considering ionization of the nitrogen- and nickel-associated defects present in the sample, and compared the calculated spectra obtained in this way to the measured spectra.

To calculate the energy dependence of defects' ionization cross-sections, we apply \emph{ab initio} Kohn-Sham DFT calculations. In the photo-ionization process, an electron is excited from an in-gap defect level to the conduction band (CB) or from the valence band (VB) to an in-gap defect level. The photoionization probability is then directly proportional to the absorption cross-section that depends on the imaginary part of the dielectric function related to the transition between the initial ground state and the final excited state. This process can be well approximated by the transition of a single electron from/to the in-gap defect level to/from the band edges, thus the imaginary part of the dielectric function can be calculated between the corresponding Kohn-Sham states and levels. Our tasks are therefore to calculate (a) the excitation energies and (b) the corresponding imaginary part of the dielectric function.

(a)	In the optical excitation process the atoms may change their geometry w.r.t.\ their ground state. We consider this effect within the Franck-Condon approximation, and calculate the lowest energy excitation that corresponds to the pure electronic transition, i.e., the zero-phonon line (ZPL) energy. This is calculated with the so called $\Delta$-self-consistent-field ($\Delta$SCF) DFT approach as was demonstrated for NV center~\cite{gali09}. Here it is critical to apply a DFT functional which is able to reproduce the band gap of diamond and the in-gap defect levels. According to our previous studies~\cite{gali09, deak}, this can be achieved by a range-separated and screened hybrid density functional (functional of Heyd-Scuseria-Ernzerhof -- HSE06)\cite{heyd, krukau}. Here we particularly calculated the ZPL energies only for the band edges explicitly, and we assumed that the larger excitation energies follow the calculated bands w.r.t.\ the band edges. We note that an alternative is to simply calculate the adiabatic charge transition level. In this case, the exciton binding energy is neglected. As shown below, the exciton binding energy can be indeed neglected.

(b)	In the Franck-Condon approximation the imaginary part of the dielectric function for the ZPL transition is the same as it is for the corresponding phonon sidebands. Thus, the imaginary part of the dielectric function can be calculated at the ground state geometry. We note that optical transitions to the bands require an accurate calculation of the electron density of states. In practice, this implies to involve many k-points in the Brillouin-zone. However, HSE06 calculation with many k-points is computationally prohibitive in a large supercell embedding the defect. Thus, we applied the semilocal Perdew-Burke-Ernzerhof (PBE) functional~\cite{perdew96} since PBE and HSE06 functionals produce very similar Kohn-Sham wavefunctions for calculating the optical transition dipole moments. 

The defect was modeled in a 512-atom diamond supercell. We applied the 
$\Gamma$-point for the geometry optimization of the defects. We note that the 
$\Gamma$-point in this supercell folds the k-point of the conduction band minimum. The geometry is converged when the forces acting on the atoms are lower than 0.02~eV/\AA. We calculate the DFT charge and spin densities of the systems by \textsc{vasp} code~\cite{VASP1, VASP2}, using the projector-augmented wave (PAW) formalism~\cite{paw}. Standard PAW potentials are chosen for the N and C ions and Ni 3d electrons are treated as valence electrons. Plane-wave basis set is utilized to expand the wave function of valence electrons. The energy cutoff for the expansion of the plane waves is set to 370~eV. The total energy of the charged supercell was corrected both in the ground and excited states to cancel the finite size errors~\cite{freysoldt1}. We found that $6\times6\times6$ $\Gamma$-centered k-point sampling of the Brillouin-zone produces converged electron density states in the conduction bands and the valence bands. We applied this k-point set for defects possessing $C_{3v}$ symmetry in the calculation of the imaginary part of the dielectric function. We calculated this property of defects with $C_{1h}$ symmetry only within $4\times4\times4$ $\Gamma$-centered k-point sampling of the Brillouin-zone, yielding also converged results in the energy region of interest. The calculated optical absorption functions were finally constructed by following the recipe of (a) and (b) with additional Gaussian smearing of 0.2~eV to simulate the vibration effects at room temperature.

To reconstruct the photocurrent spectra, the calculation was performed for 
all the major defects present in the sample in their relevant charge states. We first discuss the photophysics of NV and N$_s$ defects in detail (see Table~\ref{tab:transitions}). For N$_s$ the neutral and positive charge states are considered, whereas neutral, negative and positive charge states are taken into account for NV (see 
Ref.~\onlinecite{deak2014}). We find that the exciton binding energy of these defects (see Supplemental Material~[\onlinecite{SM}]) can be neglected at room temperature with respect to the broadening effects due to vibrations in the photocurrent spectrum.
\begin{table}[ht]
\caption{\label{tab:transitions}
Relevant ionization energies used in the construction of the photocurrent 
spectrum that are given in eV unit. The NV$^+$ is shown for the sake of 
completeness but it does not appear in the photocurrent spectrum. For NV and 
NiV defects the ionization energies are determined from HSE06 DFT calculations, 
whereas the experimental one is applied for N$_\text{s}^0$. UV means ultraviolet 
photoexcitation energy that is outside the range of our experiments.}
\begin{ruledtabular}
\begin{tabular}{lccccc}
Defect  & $(+)$$\rightarrow$$(0)$ & $(0)$$\rightarrow$$(+)$ & 
$(0)$$\rightarrow$$(-)$ &  $(-)$$\rightarrow$$(0)$ & 
$(-)$$\rightarrow$$(2-)$ \\
 \hline
  NV                  &  0.99 & UV   & 2.74 &  2.78   &         \\
  N$_\text{s}$  & UV    & 2.20 &        &            &         \\
  NiV                 &          &        & 1.22 &  UV     &  1.90  \\
   \end{tabular}
\end{ruledtabular}
\end{table}

The following conclusions can be drawn for NV defect from these results: (i) the NV$^-$$\rightarrow$NV$^0$ and NV$^0$$\rightarrow$NV$^-$ transition energies almost coincide at $\sim$2.76~eV, thus the individual ionization bands cannot be resolved in the room temperature photocurrent spectrum; (ii) the estimated energy of the intra defect level optical transition (see Ref.~\onlinecite{SM}) is larger than the band edge excitation energy for NV$^+$, thus NV$^+$ might show up as a weak absorption at around 1~eV; (iii) NV$^0$ can be ionized to NV$^+$ only by ultraviolet excitation ($\approx$4.2~eV) but NV$^0$$\rightarrow$NV$^-$ transition is active in this energy region. We conclude that photoexcitation in the visible range induces cycles between NV$^-$ and NV$^0$, and transitions to NV$^+$ do not occur. In addition, the calculated NV$^+$$\rightarrow$NV$^0$ transition energy is lower by 0.2~eV than the 1.2~eV ionization band in the photocurrent spectrum that is above the uncertainty of our method. We conclude that NV$^+$ does not show up in the photocurrent spectrum.

The photoionization of N$_\text{s}^0$ requires a short discussion. This defect 
has a giant reorganization energy (1.37~eV, see Ref.~\onlinecite{SM}) 
of the ions upon ionization. The reason behind this phenomena is that 
N$_\text{s}^0$ 
is a Jahn-Teller unstable system in $T_d$ symmetry and it strongly reconstructs 
to $C_{3v}$ symmetry. Thus, the ionization cross section is very minor at the ZPL 
energy but much stronger with the participation of phonons that drive the atoms 
from the high $T_d$ symmetry to the reconstructed $C_{3v}$ symmetry. This 
effect was already discussed in previous studies~\cite{nesladek98, Rosa99}. In 
experimental studies~\cite{nesladek98, Rosa99}, the photoionization threshold of 
N$_\text{s}$ was found at $\approx$2.2~eV by using the Inkson's 
formula~\cite{inkson81} for the 
photoionization cross section of deep level defects. This finding explains long-standing unresolved discrepancy between the thermal ionization of 
N$_\text{s}^0$, i.e., its energy in the bandgap at thermal equilibrium 
(1.7~eV below the CB), and its photoionization energy (2.2~eV). However, it is beyond the scope of this study to directly involve phonons in the calculated
photocurrent spectrum. We rather used the experimental value of 2.2~eV for the
ionization threshold of N$_\text{s}^0$ instead of the calculated ZPL energy. 

The neutral and the negative charge states of the NiV color center were already 
studied in a previous work by means of our \emph{ab initio} method~\cite{thiering}. We note that the split-vacancy configuration of this defect constitutes a high $D_{3d}$ symmetry where Ni sits at the inversion center of diamond near two adjacent vacancies, i.e., in divacancy. 
This defect may be labeled as V-Ni-V, in order
to note the interstitial position of the Ni impurity atom. However, quantum optic groups often label the split-vacancy complexes of X impurity as XV color center, e.g., group-IV -- vacancy complexes. Thus, we decided to use this nomenclature for Ni too. The detailed discussion of the electronic structure of the defect 
can be found in Ref.~\onlinecite{thiering}: briefly, a double degenerate $e_u$ level appears in the gap which is occupied by two electrons with parallel spins in the neutral charge state that establishes an $S=1$ ground state. 
This can be further occupied by one or two electrons that leads to single and 
double acceptor states and levels with $S=1/2$ and $S=0$ spin states, 
respectively (see Table~\ref{tab:transitions}). 

\begin{figure}[h]
\includegraphics[width=0.48\textwidth]{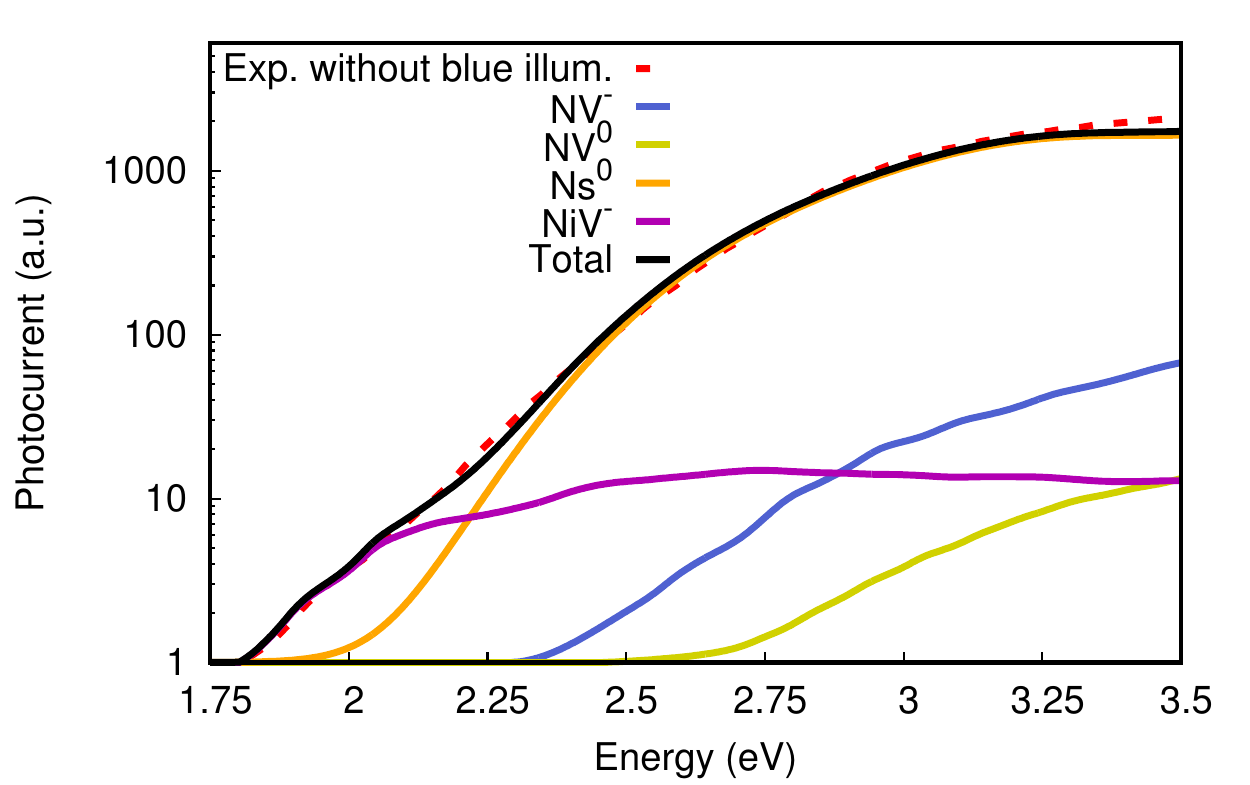}
\caption{\label{fig:noBlue}
The photocurrent spectrum measured on our sample in the absence of blue bias light (method 1) is fitted to the concentrations of NV$^-$, NV$^0$ and NiV$^-$ defects using \emph{ab initio} calculated ionization cross sections. The concentration of N$_\text{s}$ was fixed at 200~ppm. The defect concentration resulting from the fit are listed in Table~\ref{tab:conc}, in the line entitled "no blue".   
}
\end{figure}

To reconstruct the photocurrent spectrum measured following the method 1 [no blue bias light, see Fig.~\ref{fig:exp}(a)], we assumed that all N$_\text{s}$ centers are in the neutral charge state, thus we fixed the concentration of N$_\text{s}^0$ at the experimental value of 200~ppm. We then varied the concentrations of NV$^-$ and NV$^0$ (threshold around 2.76~eV) and NiV$^-$ (threshold around 1.9~eV), in order to fit to the experimental photocurrent spectrum (dotted red line in Fig.~\ref{fig:noBlue}). We note that the calculated ionization cross section in a wide region of energies is significantly larger for NV$^0$ than that for NV$^-$. This explains why green photoexcitation of individual NV defects can stabilize NV$^-$ over NV$^0$. From the fit we find the calculated $[$NV$^-]$ and $[$NV$^0]$ at 31.4~ppm and 1.0~ppm, respectively, i.e., close to the values of $\approx$34~ppm and $\approx$1.1~ppm, respectively, deduced from experiments (see Table~\ref{tab:conc}). The fitted concentration of NiV$^-$ is about 1.2~ppm. It is likely that these Ni defects reside in the region of diamond with neutral NV centers where the quasi Fermi-level is lower than that in the other regions of diamond. 

\begin{figure}
\includegraphics[width=0.48\textwidth]{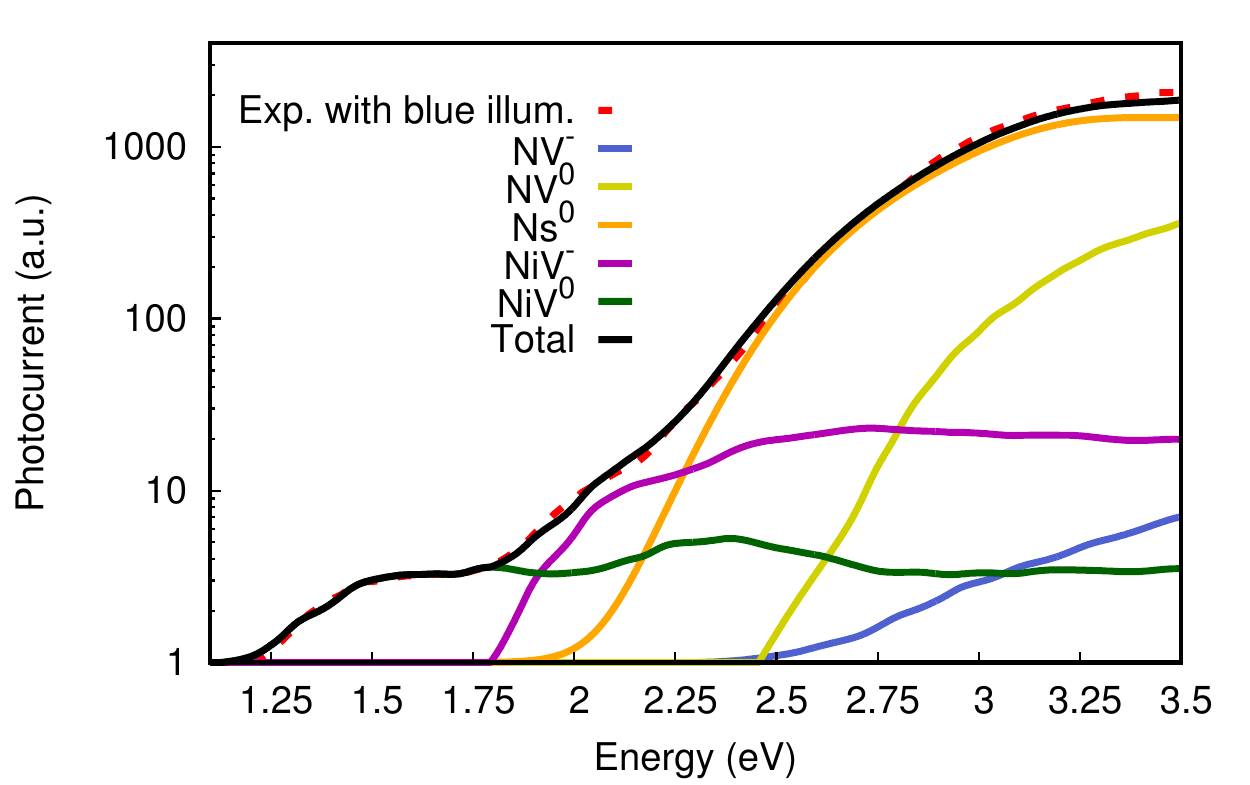}
\caption{\label{fig:withBlue}
The photocurrent spectrum measured on our sample under blue illumination (method 2) is fitted to the concentrations of N$_\text{s}^0$, NV$^-$, NV$^0$, 
NiV$^-$ and NiV$^0$ defects with using \emph{ab initio} calculated ionization cross sections. The defect concentrations resulting from the fit are listed in Table~\ref{tab:conc}, in the line entitled "blue".
}
\end{figure}

To reconstruct the photocurrent spectrum obtained under blue bias illumination [method 2, see Fig.~\ref{fig:exp}(a)], we associate the ionization band with threshold around 1.2~eV to the ionization of NiV$^0$ (see red dotted curve in Fig.~\ref{fig:withBlue}). Indeed, the results of our simulation imply that blue illumination produces holes that yield more NV$^0$, NiV$^-$ and NiV$^0$, at the expense of less N$_s^0$ and 
NV$^-$. The calculated concentration of NiV$^0$ is 0.14~ppm under blue bias illumination (see Table~\ref{tab:conc}), whereas that of NiV$^-$ increased substantially compared to that obtained by fitting the spectrum measured in the absence of bias light. We attribute this effect to the fact that NiV$^{2-}$ defects capture holes and are converted to NiV$^{-}$ and, with less extent, NiV$^{0}$ under blue illumination. The results are summarized in Table~\ref{tab:conc}. 
\begin{table}[ht]
\caption{\label{tab:conc} Concentration of defects in ppm unit as determined from 
characterization of the sample by PL and FTIR spectroscopy (noted "dark exp.") and by fitting of the photocurrent spectra measured without and with blue illumination (noted "no-blue" and "blue", respectively) using \emph{ab initio} calculated ionization cross sections. The concentration of NiV complex could not be deduced from experiments.}
\begin{ruledtabular}
\begin{tabular}{lccccc}
Sample     & [NV$^-$] & [NV$^0$] & [N$_s^0$] & [NiV$^-$] & [NiV$^0$] \\
 \hline
 dark exp.  &     34       &   1.1       &  220$\pm$20 &   &   \\
 no-blue    &  31.4        &  1.0        &  200.0          &   1.2         &     0.0          \\
 blue         &    2.9       &  29.5       &  180.0          &   2.0         &     0.14        \\
   \end{tabular}
\end{ruledtabular} 
\end{table}

Our combined experimental and atomistic simulation study identified the NiV 
defect in an irradiated and annealed type-Ib diamond sample. We note 
that an absorption center was found at 1.22~eV in Ni contaminated 
diamond~\cite{lawson1993} that was correlated with the NOL1/NIRIM5 electron 
paramagnetic resonance (EPR) spectrum~\cite{nadolinny2002, Iakoubovskii04}. 
Based on our modeling, the 1.22-eV absorption peak can be well explained by the 
optical transition from the valence band edges to the empty defect level in the 
spin minority channel. The NOL1/NIRIM5 EPR center possesses a giant 
zero-field-splitting of $D=-171$~GHz, and a relatively large anisotropy in the $g$-tensor, 
$g_\|$=2.0235, and $g_\perp$=2.0020~\cite{nadolinny2002, Iakoubovskii04}. 
This large anisotropy implies a strong second order contribution of the spin-orbit 
interaction in both the $g$-tensor and the $D$-tensor. Indeed, we find only 
$\approx$0.7~GHz contribution to the zero-field-splitting from the electron spin-
electron spin -- dipole-dipole interaction (see Ref.~\onlinecite{ivady2014} for the 
method) that supports this model. Furthermore, the calculated hyperfine 
constants (see Ref.~\onlinecite{szasz2013} for the method) of the $^{61}$Ni 
are smaller ($A_\perp$=25~MHz; $A_\|$=63~MHz) than those of $^{13}$C  
($A_\perp$=39~MHz; $A_\|$=84~MHz) of the 6 nearest neighbor C atom of the 
Ni atom in diamond, which agrees well with the NOL1/NIRIM5 EPR spectrum (see Fig.\ 2 in Ref.~\onlinecite{Iakoubovskii04}). This confirms that the NOL1/NIRIM5
EPR center, that is identified as NiV$^0$, has a stable $S=1$ spin state at T=20~K 
temperature~\cite{Iakoubovskii04} and that optically active excited states exist. 
Similarly to the neutral silicon-vacancy center~\cite{edmonds08}, this defect is 
thermally stable in boron doped diamond~\cite{Iakoubovskii04}, in agreement 
with our model. Once the optical spinpolarization is confirmed for NiV$^0$ then 
our photocurrent spectroscopy measurements show that this center shall be a very promising candidate for photoelectric detection of the spin state operating in the near-infrared region for photoionization.

In this paper, we recorded the photocurrent spectrum of diamond samples that 
contain nitrogen and nickel impurities. We characterized the photophysics of NV 
center and substitutional nitrogen defect in diamond by means of first principles 
calculations. We identified the $\sim$1.2 and $\sim$1.9~eV photoionization bands
observed by photocurrent spectroscopy with the single and double acceptor levels of the nickel split vacancy complex NiV. Ionization of the NiV defect contributes therefore to the total photocurrent measured under the green illumination used in the PDMR technique to induce the two-photon ionization of NV centers. Our calculations support that the NOL1/NIRIM5 EPR center and the 1.22-eV absorption center originate from the $S=1$ neutral NiV defect that could potentially be used as a photoelectrically readout solid state qubit operating at RT and excited by near-infrared light, presenting interest for quantum information technology.

This work was supported by EU Commission (project DIADEMS with contract No.~611143). A.G.\ acknowledges the support from National 
Research Development and Innovation Office of Hungary within
the Quantum Technology National Excellence Program (project no.
2017-1.2.1-NKP-2017-00001) and QuantERA Q-Magine project (contract no. 127889).


\begin{thebibliography}{49}%
\makeatletter
\providecommand \@ifxundefined [1]{%
 \@ifx{#1\undefined}
}%
\providecommand \@ifnum [1]{%
 \ifnum #1\expandafter \@firstoftwo
 \else \expandafter \@secondoftwo
 \fi
}%
\providecommand \@ifx [1]{%
 \ifx #1\expandafter \@firstoftwo
 \else \expandafter \@secondoftwo
 \fi
}%
\providecommand \natexlab [1]{#1}%
\providecommand \enquote  [1]{``#1''}%
\providecommand \bibnamefont  [1]{#1}%
\providecommand \bibfnamefont [1]{#1}%
\providecommand \citenamefont [1]{#1}%
\providecommand \href@noop [0]{\@secondoftwo}%
\providecommand \href [0]{\begingroup \@sanitize@url \@href}%
\providecommand \@href[1]{\@@startlink{#1}\@@href}%
\providecommand \@@href[1]{\endgroup#1\@@endlink}%
\providecommand \@sanitize@url [0]{\catcode `\\12\catcode `\$12\catcode
  `\&12\catcode `\#12\catcode `\^12\catcode `\_12\catcode `\%12\relax}%
\providecommand \@@startlink[1]{}%
\providecommand \@@endlink[0]{}%
\providecommand \url  [0]{\begingroup\@sanitize@url \@url }%
\providecommand \@url [1]{\endgroup\@href {#1}{\urlprefix }}%
\providecommand \urlprefix  [0]{URL }%
\providecommand \Eprint [0]{\href }%
\providecommand \doibase [0]{http://dx.doi.org/}%
\providecommand \selectlanguage [0]{\@gobble}%
\providecommand \bibinfo  [0]{\@secondoftwo}%
\providecommand \bibfield  [0]{\@secondoftwo}%
\providecommand \translation [1]{[#1]}%
\providecommand \BibitemOpen [0]{}%
\providecommand \bibitemStop [0]{}%
\providecommand \bibitemNoStop [0]{.\EOS\space}%
\providecommand \EOS [0]{\spacefactor3000\relax}%
\providecommand \BibitemShut  [1]{\csname bibitem#1\endcsname}%
\let\auto@bib@innerbib\@empty
\bibitem [{\citenamefont {Zaitsev}(2001)}]{Zai01}%
  \BibitemOpen
  \bibfield  {author} {\bibinfo {author} {\bibfnamefont {A.~M.}\ \bibnamefont
  {Zaitsev}},\ }\href {http://link.springer.com/10.1007/978-3-662-04548-0}
  {\emph {\bibinfo {title} {Optical Properties of Diamond}}}\ (\bibinfo
  {publisher} {Springer Berlin Heidelberg},\ \bibinfo {address} {Berlin,
  Heidelberg},\ \bibinfo {year} {2001})\ \bibinfo {note} {dOI:
  10.1007/978-3-662-04548-0}\BibitemShut {NoStop}%
\bibitem [{\citenamefont {Jelezko}\ and\ \citenamefont
  {Wrachtrup}(2006)}]{jel06}%
  \BibitemOpen
  \bibfield  {author} {\bibinfo {author} {\bibfnamefont {F.}~\bibnamefont
  {Jelezko}}\ and\ \bibinfo {author} {\bibfnamefont {J.}~\bibnamefont
  {Wrachtrup}},\ }\href {\doibase 10.1002/pssa.200671403} {\bibfield  {journal}
  {\bibinfo  {journal} {physica status solidi (a)}\ }\textbf {\bibinfo {volume}
  {203}},\ \bibinfo {pages} {3207} (\bibinfo {year} {2006})}\BibitemShut
  {NoStop}%
\bibitem [{\citenamefont {Childress}\ \emph {et~al.}(2006)\citenamefont
  {Childress}, \citenamefont {Gurudev~Dutt}, \citenamefont {Taylor},
  \citenamefont {Zibrov}, \citenamefont {Jelezko}, \citenamefont {Wrachtrup},
  \citenamefont {Hemmer},\ and\ \citenamefont {Lukin}}]{childress06}%
  \BibitemOpen
  \bibfield  {author} {\bibinfo {author} {\bibfnamefont {L.}~\bibnamefont
  {Childress}}, \bibinfo {author} {\bibfnamefont {M.~V.}\ \bibnamefont
  {Gurudev~Dutt}}, \bibinfo {author} {\bibfnamefont {J.~M.}\ \bibnamefont
  {Taylor}}, \bibinfo {author} {\bibfnamefont {A.~S.}\ \bibnamefont {Zibrov}},
  \bibinfo {author} {\bibfnamefont {F.}~\bibnamefont {Jelezko}}, \bibinfo
  {author} {\bibfnamefont {J.}~\bibnamefont {Wrachtrup}}, \bibinfo {author}
  {\bibfnamefont {P.~R.}\ \bibnamefont {Hemmer}}, \ and\ \bibinfo {author}
  {\bibfnamefont {M.~D.}\ \bibnamefont {Lukin}},\ }\href {\doibase
  10.1126/science.1131871} {\bibfield  {journal} {\bibinfo  {journal}
  {Science}\ }\textbf {\bibinfo {volume} {314}},\ \bibinfo {pages} {281}
  (\bibinfo {year} {2006})},\ \Eprint
  {http://arxiv.org/abs/http://science.sciencemag.org/content/314/5797/281.full.pdf}
  {http://science.sciencemag.org/content/314/5797/281.full.pdf} \BibitemShut
  {NoStop}%
\bibitem [{\citenamefont {Maze}\ \emph {et~al.}(2008)\citenamefont {Maze},
  \citenamefont {Stanwix}, \citenamefont {Hodges}, \citenamefont {Hong},
  \citenamefont {Taylor}, \citenamefont {Cappellaro}, \citenamefont {Jiang},
  \citenamefont {Dutt}, \citenamefont {Togan}, \citenamefont {Zibrov},
  \citenamefont {Yacoby}, \citenamefont {Walsworth},\ and\ \citenamefont
  {Lukin}}]{maze08}%
  \BibitemOpen
  \bibfield  {author} {\bibinfo {author} {\bibfnamefont {J.~R.}\ \bibnamefont
  {Maze}}, \bibinfo {author} {\bibfnamefont {P.~L.}\ \bibnamefont {Stanwix}},
  \bibinfo {author} {\bibfnamefont {J.~S.}\ \bibnamefont {Hodges}}, \bibinfo
  {author} {\bibfnamefont {S.}~\bibnamefont {Hong}}, \bibinfo {author}
  {\bibfnamefont {J.~M.}\ \bibnamefont {Taylor}}, \bibinfo {author}
  {\bibfnamefont {P.}~\bibnamefont {Cappellaro}}, \bibinfo {author}
  {\bibfnamefont {L.}~\bibnamefont {Jiang}}, \bibinfo {author} {\bibfnamefont
  {M.~V.~G.}\ \bibnamefont {Dutt}}, \bibinfo {author} {\bibfnamefont
  {E.}~\bibnamefont {Togan}}, \bibinfo {author} {\bibfnamefont {A.~S.}\
  \bibnamefont {Zibrov}}, \bibinfo {author} {\bibfnamefont {A.}~\bibnamefont
  {Yacoby}}, \bibinfo {author} {\bibfnamefont {R.~L.}\ \bibnamefont
  {Walsworth}}, \ and\ \bibinfo {author} {\bibfnamefont {M.~D.}\ \bibnamefont
  {Lukin}},\ }\href {\doibase 10.1038/nature07279} {\bibfield  {journal}
  {\bibinfo  {journal} {Nature}\ }\textbf {\bibinfo {volume} {455}},\ \bibinfo
  {pages} {644} (\bibinfo {year} {2008})}\BibitemShut {NoStop}%
\bibitem [{\citenamefont {Balasubramanian}\ \emph {et~al.}(2008)\citenamefont
  {Balasubramanian}, \citenamefont {Chan}, \citenamefont {Kolesov},
  \citenamefont {Al-Hmoud}, \citenamefont {Tisler}, \citenamefont {Shin},
  \citenamefont {Kim}, \citenamefont {Wojcik}, \citenamefont {Hemmer},
  \citenamefont {Krueger}, \citenamefont {Hanke}, \citenamefont
  {Leitenstorfer}, \citenamefont {Bratschitsch}, \citenamefont {Jelezko},\ and\
  \citenamefont {Wrachtrup}}]{balasubramanian08}%
  \BibitemOpen
  \bibfield  {author} {\bibinfo {author} {\bibfnamefont {G.}~\bibnamefont
  {Balasubramanian}}, \bibinfo {author} {\bibfnamefont {I.~Y.}\ \bibnamefont
  {Chan}}, \bibinfo {author} {\bibfnamefont {R.}~\bibnamefont {Kolesov}},
  \bibinfo {author} {\bibfnamefont {M.}~\bibnamefont {Al-Hmoud}}, \bibinfo
  {author} {\bibfnamefont {J.}~\bibnamefont {Tisler}}, \bibinfo {author}
  {\bibfnamefont {C.}~\bibnamefont {Shin}}, \bibinfo {author} {\bibfnamefont
  {C.}~\bibnamefont {Kim}}, \bibinfo {author} {\bibfnamefont {A.}~\bibnamefont
  {Wojcik}}, \bibinfo {author} {\bibfnamefont {P.~R.}\ \bibnamefont {Hemmer}},
  \bibinfo {author} {\bibfnamefont {A.}~\bibnamefont {Krueger}}, \bibinfo
  {author} {\bibfnamefont {T.}~\bibnamefont {Hanke}}, \bibinfo {author}
  {\bibfnamefont {A.}~\bibnamefont {Leitenstorfer}}, \bibinfo {author}
  {\bibfnamefont {R.}~\bibnamefont {Bratschitsch}}, \bibinfo {author}
  {\bibfnamefont {F.}~\bibnamefont {Jelezko}}, \ and\ \bibinfo {author}
  {\bibfnamefont {J.}~\bibnamefont {Wrachtrup}},\ }\href {\doibase
  10.1038/nature07278} {\bibfield  {journal} {\bibinfo  {journal} {Nature}\
  }\textbf {\bibinfo {volume} {455}},\ \bibinfo {pages} {648} (\bibinfo {year}
  {2008})}\BibitemShut {NoStop}%
\bibitem [{\citenamefont {Rondin}\ \emph {et~al.}(2014)\citenamefont {Rondin},
  \citenamefont {Tetienne}, \citenamefont {Hingant}, \citenamefont {Roch},
  \citenamefont {Maletinsky},\ and\ \citenamefont {Jacques}}]{rondin14}%
  \BibitemOpen
  \bibfield  {author} {\bibinfo {author} {\bibfnamefont {L.}~\bibnamefont
  {Rondin}}, \bibinfo {author} {\bibfnamefont {J.-P.}\ \bibnamefont
  {Tetienne}}, \bibinfo {author} {\bibfnamefont {T.}~\bibnamefont {Hingant}},
  \bibinfo {author} {\bibfnamefont {J.-F.}\ \bibnamefont {Roch}}, \bibinfo
  {author} {\bibfnamefont {P.}~\bibnamefont {Maletinsky}}, \ and\ \bibinfo
  {author} {\bibfnamefont {V.}~\bibnamefont {Jacques}},\ }\href@noop {}
  {\bibfield  {journal} {\bibinfo  {journal} {Reports on Progress in Physics}\
  }\textbf {\bibinfo {volume} {77}},\ \bibinfo {pages} {056503} (\bibinfo
  {year} {2014})}\BibitemShut {NoStop}%
\bibitem [{\citenamefont {Dolde}\ \emph {et~al.}(2014)\citenamefont {Dolde},
  \citenamefont {Doherty}, \citenamefont {Michl}, \citenamefont {Jakobi},
  \citenamefont {Naydenov}, \citenamefont {Pezzagna}, \citenamefont {Meijer},
  \citenamefont {Neumann}, \citenamefont {Jelezko}, \citenamefont {Manson},\
  and\ \citenamefont {Wrachtrup}}]{dolde14}%
  \BibitemOpen
  \bibfield  {author} {\bibinfo {author} {\bibfnamefont {F.}~\bibnamefont
  {Dolde}}, \bibinfo {author} {\bibfnamefont {M.~W.}\ \bibnamefont {Doherty}},
  \bibinfo {author} {\bibfnamefont {J.}~\bibnamefont {Michl}}, \bibinfo
  {author} {\bibfnamefont {I.}~\bibnamefont {Jakobi}}, \bibinfo {author}
  {\bibfnamefont {B.}~\bibnamefont {Naydenov}}, \bibinfo {author}
  {\bibfnamefont {S.}~\bibnamefont {Pezzagna}}, \bibinfo {author}
  {\bibfnamefont {J.}~\bibnamefont {Meijer}}, \bibinfo {author} {\bibfnamefont
  {P.}~\bibnamefont {Neumann}}, \bibinfo {author} {\bibfnamefont
  {F.}~\bibnamefont {Jelezko}}, \bibinfo {author} {\bibfnamefont {N.~B.}\
  \bibnamefont {Manson}}, \ and\ \bibinfo {author} {\bibfnamefont
  {J.}~\bibnamefont {Wrachtrup}},\ }\href {\doibase
  10.1103/PhysRevLett.112.097603} {\bibfield  {journal} {\bibinfo  {journal}
  {Phys. Rev. Lett.}\ }\textbf {\bibinfo {volume} {112}},\ \bibinfo {pages}
  {097603} (\bibinfo {year} {2014})}\BibitemShut {NoStop}%
\bibitem [{\citenamefont {Neumann}\ \emph {et~al.}(2013)\citenamefont
  {Neumann}, \citenamefont {Jakobi}, \citenamefont {Dolde}, \citenamefont
  {Burk}, \citenamefont {Reuter}, \citenamefont {Waldherr}, \citenamefont
  {Honert}, \citenamefont {Wolf}, \citenamefont {Brunner}, \citenamefont
  {Shim}, \citenamefont {Suter}, \citenamefont {Sumiya}, \citenamefont
  {Isoya},\ and\ \citenamefont {Wrachtrup}}]{neumann13}%
  \BibitemOpen
  \bibfield  {author} {\bibinfo {author} {\bibfnamefont {P.}~\bibnamefont
  {Neumann}}, \bibinfo {author} {\bibfnamefont {I.}~\bibnamefont {Jakobi}},
  \bibinfo {author} {\bibfnamefont {F.}~\bibnamefont {Dolde}}, \bibinfo
  {author} {\bibfnamefont {C.}~\bibnamefont {Burk}}, \bibinfo {author}
  {\bibfnamefont {R.}~\bibnamefont {Reuter}}, \bibinfo {author} {\bibfnamefont
  {G.}~\bibnamefont {Waldherr}}, \bibinfo {author} {\bibfnamefont
  {J.}~\bibnamefont {Honert}}, \bibinfo {author} {\bibfnamefont
  {T.}~\bibnamefont {Wolf}}, \bibinfo {author} {\bibfnamefont {A.}~\bibnamefont
  {Brunner}}, \bibinfo {author} {\bibfnamefont {J.~H.}\ \bibnamefont {Shim}},
  \bibinfo {author} {\bibfnamefont {D.}~\bibnamefont {Suter}}, \bibinfo
  {author} {\bibfnamefont {H.}~\bibnamefont {Sumiya}}, \bibinfo {author}
  {\bibfnamefont {J.}~\bibnamefont {Isoya}}, \ and\ \bibinfo {author}
  {\bibfnamefont {J.}~\bibnamefont {Wrachtrup}},\ }\href {\doibase
  10.1021/nl401216y} {\bibfield  {journal} {\bibinfo  {journal} {Nano Letters}\
  }\textbf {\bibinfo {volume} {13}},\ \bibinfo {pages} {2738} (\bibinfo {year}
  {2013})},\ \bibinfo {note} {pMID: 23721106}\BibitemShut {NoStop}%
\bibitem [{\citenamefont {Toyli}\ \emph {et~al.}(2013)\citenamefont {Toyli},
  \citenamefont {de~las Casas}, \citenamefont {Christle}, \citenamefont
  {Dobrovitski},\ and\ \citenamefont {Awschalom}}]{toyli13}%
  \BibitemOpen
  \bibfield  {author} {\bibinfo {author} {\bibfnamefont {D.~M.}\ \bibnamefont
  {Toyli}}, \bibinfo {author} {\bibfnamefont {C.~F.}\ \bibnamefont {de~las
  Casas}}, \bibinfo {author} {\bibfnamefont {D.~J.}\ \bibnamefont {Christle}},
  \bibinfo {author} {\bibfnamefont {V.~V.}\ \bibnamefont {Dobrovitski}}, \ and\
  \bibinfo {author} {\bibfnamefont {D.~D.}\ \bibnamefont {Awschalom}},\ }\href
  {\doibase 10.1073/pnas.1306825110} {\bibfield  {journal} {\bibinfo  {journal}
  {Proceedings of the National Academy of Sciences}\ }\textbf {\bibinfo
  {volume} {110}},\ \bibinfo {pages} {8417} (\bibinfo {year}
  {2013})}\BibitemShut {NoStop}%
\bibitem [{\citenamefont {Maurer}\ \emph {et~al.}(2012)\citenamefont {Maurer},
  \citenamefont {Kucsko}, \citenamefont {Latta}, \citenamefont {Jiang},
  \citenamefont {Yao}, \citenamefont {Bennett}, \citenamefont {Pastawski},
  \citenamefont {Hunger}, \citenamefont {Chisholm}, \citenamefont {Markham},
  \citenamefont {Twitchen}, \citenamefont {Cirac},\ and\ \citenamefont
  {Lukin}}]{maurer12}%
  \BibitemOpen
  \bibfield  {author} {\bibinfo {author} {\bibfnamefont {P.~C.}\ \bibnamefont
  {Maurer}}, \bibinfo {author} {\bibfnamefont {G.}~\bibnamefont {Kucsko}},
  \bibinfo {author} {\bibfnamefont {C.}~\bibnamefont {Latta}}, \bibinfo
  {author} {\bibfnamefont {L.}~\bibnamefont {Jiang}}, \bibinfo {author}
  {\bibfnamefont {N.~Y.}\ \bibnamefont {Yao}}, \bibinfo {author} {\bibfnamefont
  {S.~D.}\ \bibnamefont {Bennett}}, \bibinfo {author} {\bibfnamefont
  {F.}~\bibnamefont {Pastawski}}, \bibinfo {author} {\bibfnamefont
  {D.}~\bibnamefont {Hunger}}, \bibinfo {author} {\bibfnamefont
  {N.}~\bibnamefont {Chisholm}}, \bibinfo {author} {\bibfnamefont
  {M.}~\bibnamefont {Markham}}, \bibinfo {author} {\bibfnamefont {D.~J.}\
  \bibnamefont {Twitchen}}, \bibinfo {author} {\bibfnamefont {J.~I.}\
  \bibnamefont {Cirac}}, \ and\ \bibinfo {author} {\bibfnamefont {M.~D.}\
  \bibnamefont {Lukin}},\ }\href {\doibase 10.1126/science.1220513} {\bibfield
  {journal} {\bibinfo  {journal} {Science}\ }\textbf {\bibinfo {volume}
  {336}},\ \bibinfo {pages} {1283} (\bibinfo {year} {2012})}\BibitemShut
  {NoStop}%
\bibitem [{\citenamefont {Pfaff}\ \emph {et~al.}(2014)\citenamefont {Pfaff},
  \citenamefont {Hensen}, \citenamefont {Bernien}, \citenamefont {van Dam},
  \citenamefont {Blok}, \citenamefont {Taminiau}, \citenamefont {Tiggelman},
  \citenamefont {Schouten}, \citenamefont {Markham}, \citenamefont {Twitchen},\
  and\ \citenamefont {Hanson}}]{pfaff14}%
  \BibitemOpen
  \bibfield  {author} {\bibinfo {author} {\bibfnamefont {W.}~\bibnamefont
  {Pfaff}}, \bibinfo {author} {\bibfnamefont {B.~J.}\ \bibnamefont {Hensen}},
  \bibinfo {author} {\bibfnamefont {H.}~\bibnamefont {Bernien}}, \bibinfo
  {author} {\bibfnamefont {S.~B.}\ \bibnamefont {van Dam}}, \bibinfo {author}
  {\bibfnamefont {M.~S.}\ \bibnamefont {Blok}}, \bibinfo {author}
  {\bibfnamefont {T.~H.}\ \bibnamefont {Taminiau}}, \bibinfo {author}
  {\bibfnamefont {M.~J.}\ \bibnamefont {Tiggelman}}, \bibinfo {author}
  {\bibfnamefont {R.~N.}\ \bibnamefont {Schouten}}, \bibinfo {author}
  {\bibfnamefont {M.}~\bibnamefont {Markham}}, \bibinfo {author} {\bibfnamefont
  {D.~J.}\ \bibnamefont {Twitchen}}, \ and\ \bibinfo {author} {\bibfnamefont
  {R.}~\bibnamefont {Hanson}},\ }\href {\doibase 10.1126/science.1253512}
  {\bibfield  {journal} {\bibinfo  {journal} {Science}\ }\textbf {\bibinfo
  {volume} {345}},\ \bibinfo {pages} {532} (\bibinfo {year}
  {2014})}\BibitemShut {NoStop}%
\bibitem [{\citenamefont {Hensen}\ \emph {et~al.}(2015)\citenamefont {Hensen},
  \citenamefont {Bernien}, \citenamefont {DrÃ©au}, \citenamefont {Reiserer},
  \citenamefont {Kalb}, \citenamefont {Blok}, \citenamefont {Ruitenberg},
  \citenamefont {Vermeulen}, \citenamefont {Schouten}, \citenamefont
  {AbellÃ¡n}, \citenamefont {Amaya}, \citenamefont {Pruneri}, \citenamefont
  {Mitchell}, \citenamefont {Markham}, \citenamefont {Twitchen}, \citenamefont
  {Elkouss}, \citenamefont {Wehner}, \citenamefont {Taminiau},\ and\
  \citenamefont {Hanson}}]{hensen15}%
  \BibitemOpen
  \bibfield  {author} {\bibinfo {author} {\bibfnamefont {B.}~\bibnamefont
  {Hensen}}, \bibinfo {author} {\bibfnamefont {H.}~\bibnamefont {Bernien}},
  \bibinfo {author} {\bibfnamefont {A.~E.}\ \bibnamefont {DrÃ©au}}, \bibinfo
  {author} {\bibfnamefont {A.}~\bibnamefont {Reiserer}}, \bibinfo {author}
  {\bibfnamefont {N.}~\bibnamefont {Kalb}}, \bibinfo {author} {\bibfnamefont
  {M.~S.}\ \bibnamefont {Blok}}, \bibinfo {author} {\bibfnamefont
  {J.}~\bibnamefont {Ruitenberg}}, \bibinfo {author} {\bibfnamefont {R.~F.~L.}\
  \bibnamefont {Vermeulen}}, \bibinfo {author} {\bibfnamefont {R.~N.}\
  \bibnamefont {Schouten}}, \bibinfo {author} {\bibfnamefont {C.}~\bibnamefont
  {AbellÃ¡n}}, \bibinfo {author} {\bibfnamefont {W.}~\bibnamefont {Amaya}},
  \bibinfo {author} {\bibfnamefont {V.}~\bibnamefont {Pruneri}}, \bibinfo
  {author} {\bibfnamefont {M.~W.}\ \bibnamefont {Mitchell}}, \bibinfo {author}
  {\bibfnamefont {M.}~\bibnamefont {Markham}}, \bibinfo {author} {\bibfnamefont
  {D.~J.}\ \bibnamefont {Twitchen}}, \bibinfo {author} {\bibfnamefont
  {D.}~\bibnamefont {Elkouss}}, \bibinfo {author} {\bibfnamefont
  {S.}~\bibnamefont {Wehner}}, \bibinfo {author} {\bibfnamefont {T.~H.}\
  \bibnamefont {Taminiau}}, \ and\ \bibinfo {author} {\bibfnamefont
  {R.}~\bibnamefont {Hanson}},\ }\href {\doibase 10.1038/nature15759}
  {\bibfield  {journal} {\bibinfo  {journal} {Nature}\ }\textbf {\bibinfo
  {volume} {526}},\ \bibinfo {pages} {682} (\bibinfo {year}
  {2015})}\BibitemShut {NoStop}%
\bibitem [{\citenamefont {Balasubramanian}\ \emph {et~al.}(2009)\citenamefont
  {Balasubramanian}, \citenamefont {Neumann}, \citenamefont {Twitchen},
  \citenamefont {Markham}, \citenamefont {Kolesov}, \citenamefont {Mizuochi},
  \citenamefont {Isoya}, \citenamefont {Achard}, \citenamefont {Beck},
  \citenamefont {Tissler}, \citenamefont {Jacques}, \citenamefont {Hemmer},
  \citenamefont {Jelezko},\ and\ \citenamefont {Wrachtrup}}]{bala09}%
  \BibitemOpen
  \bibfield  {author} {\bibinfo {author} {\bibfnamefont {G.}~\bibnamefont
  {Balasubramanian}}, \bibinfo {author} {\bibfnamefont {P.}~\bibnamefont
  {Neumann}}, \bibinfo {author} {\bibfnamefont {D.}~\bibnamefont {Twitchen}},
  \bibinfo {author} {\bibfnamefont {M.}~\bibnamefont {Markham}}, \bibinfo
  {author} {\bibfnamefont {R.}~\bibnamefont {Kolesov}}, \bibinfo {author}
  {\bibfnamefont {N.}~\bibnamefont {Mizuochi}}, \bibinfo {author}
  {\bibfnamefont {J.}~\bibnamefont {Isoya}}, \bibinfo {author} {\bibfnamefont
  {J.}~\bibnamefont {Achard}}, \bibinfo {author} {\bibfnamefont
  {J.}~\bibnamefont {Beck}}, \bibinfo {author} {\bibfnamefont {J.}~\bibnamefont
  {Tissler}}, \bibinfo {author} {\bibfnamefont {V.}~\bibnamefont {Jacques}},
  \bibinfo {author} {\bibfnamefont {P.~R.}\ \bibnamefont {Hemmer}}, \bibinfo
  {author} {\bibfnamefont {F.}~\bibnamefont {Jelezko}}, \ and\ \bibinfo
  {author} {\bibfnamefont {J.}~\bibnamefont {Wrachtrup}},\ }\href {\doibase
  10.1038/nmat2420} {\bibfield  {journal} {\bibinfo  {journal} {Nat Mater}\
  }\textbf {\bibinfo {volume} {8}},\ \bibinfo {pages} {383} (\bibinfo {year}
  {2009})}\BibitemShut {NoStop}%
\bibitem [{\citenamefont {Gruber}\ \emph {et~al.}(1997)\citenamefont {Gruber},
  \citenamefont {Dr\"abenstedt}, \citenamefont {Tietz}, \citenamefont {Fleury},
  \citenamefont {Wrachtrup},\ and\ \citenamefont {von
  Borczyskowski}}]{gruber1997}%
  \BibitemOpen
  \bibfield  {author} {\bibinfo {author} {\bibfnamefont {A.}~\bibnamefont
  {Gruber}}, \bibinfo {author} {\bibfnamefont {A.}~\bibnamefont
  {Dr\"abenstedt}}, \bibinfo {author} {\bibfnamefont {C.}~\bibnamefont
  {Tietz}}, \bibinfo {author} {\bibfnamefont {L.}~\bibnamefont {Fleury}},
  \bibinfo {author} {\bibfnamefont {J.}~\bibnamefont {Wrachtrup}}, \ and\
  \bibinfo {author} {\bibfnamefont {C.}~\bibnamefont {von Borczyskowski}},\
  }\href {\doibase 10.1126/science.276.5321.2012} {\bibfield  {journal}
  {\bibinfo  {journal} {Science}\ }\textbf {\bibinfo {volume} {276}},\ \bibinfo
  {pages} {2012} (\bibinfo {year} {1997})}\BibitemShut {NoStop}%
\bibitem [{\citenamefont {Bourgeois}\ \emph {et~al.}(2015)\citenamefont
  {Bourgeois}, \citenamefont {Jarmola}, \citenamefont {Siyushev}, \citenamefont
  {Gulka}, \citenamefont {Hruby}, \citenamefont {Jelezko}, \citenamefont
  {Budker},\ and\ \citenamefont {Nesladek}}]{bourgeois2015}%
  \BibitemOpen
  \bibfield  {author} {\bibinfo {author} {\bibfnamefont {E.}~\bibnamefont
  {Bourgeois}}, \bibinfo {author} {\bibfnamefont {A.}~\bibnamefont {Jarmola}},
  \bibinfo {author} {\bibfnamefont {P.}~\bibnamefont {Siyushev}}, \bibinfo
  {author} {\bibfnamefont {M.}~\bibnamefont {Gulka}}, \bibinfo {author}
  {\bibfnamefont {J.}~\bibnamefont {Hruby}}, \bibinfo {author} {\bibfnamefont
  {F.}~\bibnamefont {Jelezko}}, \bibinfo {author} {\bibfnamefont
  {D.}~\bibnamefont {Budker}}, \ and\ \bibinfo {author} {\bibfnamefont
  {M.}~\bibnamefont {Nesladek}},\ }\href {\doibase 10.1038/ncomms9577}
  {\bibfield  {journal} {\bibinfo  {journal} {Nature Communications}\ }\textbf
  {\bibinfo {volume} {6}},\ \bibinfo {pages} {8577} (\bibinfo {year}
  {2015})}\BibitemShut {NoStop}%
\bibitem [{\citenamefont {Hrubesch}\ \emph {et~al.}(2017)\citenamefont
  {Hrubesch}, \citenamefont {Braunbeck}, \citenamefont {Stutzmann},
  \citenamefont {Reinhard},\ and\ \citenamefont {Brandt}}]{brandt2016}%
  \BibitemOpen
  \bibfield  {author} {\bibinfo {author} {\bibfnamefont {F.~M.}\ \bibnamefont
  {Hrubesch}}, \bibinfo {author} {\bibfnamefont {G.}~\bibnamefont {Braunbeck}},
  \bibinfo {author} {\bibfnamefont {M.}~\bibnamefont {Stutzmann}}, \bibinfo
  {author} {\bibfnamefont {F.}~\bibnamefont {Reinhard}}, \ and\ \bibinfo
  {author} {\bibfnamefont {M.~S.}\ \bibnamefont {Brandt}},\ }\href {\doibase
  10.1103/PhysRevLett.118.037601} {\bibfield  {journal} {\bibinfo  {journal}
  {Phys. Rev. Lett.}\ }\textbf {\bibinfo {volume} {118}},\ \bibinfo {pages}
  {037601} (\bibinfo {year} {2017})}\BibitemShut {NoStop}%
\bibitem [{\citenamefont {Bourgeois}\ \emph {et~al.}(2017)\citenamefont
  {Bourgeois}, \citenamefont {Londero}, \citenamefont {Buczak}, \citenamefont
  {Hruby}, \citenamefont {Gulka}, \citenamefont {Balasubramaniam},
  \citenamefont {Wachter}, \citenamefont {Stursa}, \citenamefont {Dobes},
  \citenamefont {Aumayr}, \citenamefont {Trupke}, \citenamefont {Gali},\ and\
  \citenamefont {Nesladek}}]{bourgeois2016}%
  \BibitemOpen
  \bibfield  {author} {\bibinfo {author} {\bibfnamefont {E.}~\bibnamefont
  {Bourgeois}}, \bibinfo {author} {\bibfnamefont {E.}~\bibnamefont {Londero}},
  \bibinfo {author} {\bibfnamefont {K.}~\bibnamefont {Buczak}}, \bibinfo
  {author} {\bibfnamefont {J.}~\bibnamefont {Hruby}}, \bibinfo {author}
  {\bibfnamefont {M.}~\bibnamefont {Gulka}}, \bibinfo {author} {\bibfnamefont
  {Y.}~\bibnamefont {Balasubramaniam}}, \bibinfo {author} {\bibfnamefont
  {G.}~\bibnamefont {Wachter}}, \bibinfo {author} {\bibfnamefont
  {J.}~\bibnamefont {Stursa}}, \bibinfo {author} {\bibfnamefont
  {K.}~\bibnamefont {Dobes}}, \bibinfo {author} {\bibfnamefont
  {F.}~\bibnamefont {Aumayr}}, \bibinfo {author} {\bibfnamefont
  {M.}~\bibnamefont {Trupke}}, \bibinfo {author} {\bibfnamefont
  {A.}~\bibnamefont {Gali}}, \ and\ \bibinfo {author} {\bibfnamefont
  {M.}~\bibnamefont {Nesladek}},\ }\href {\doibase 10.1103/PhysRevB.95.041402}
  {\bibfield  {journal} {\bibinfo  {journal} {Phys. Rev. B}\ }\textbf {\bibinfo
  {volume} {95}},\ \bibinfo {pages} {041402} (\bibinfo {year}
  {2017})}\BibitemShut {NoStop}%
\bibitem [{\citenamefont {Gulka}\ \emph {et~al.}(2017)\citenamefont {Gulka},
  \citenamefont {Bourgeois}, \citenamefont {Hruby}, \citenamefont {Siyushev},
  \citenamefont {Wachter}, \citenamefont {Aumayr}, \citenamefont {Hemmer},
  \citenamefont {Gali}, \citenamefont {Jelezko}, \citenamefont {Trupke},\ and\
  \citenamefont {Nesladek}}]{gulka2017}%
  \BibitemOpen
  \bibfield  {author} {\bibinfo {author} {\bibfnamefont {M.}~\bibnamefont
  {Gulka}}, \bibinfo {author} {\bibfnamefont {E.}~\bibnamefont {Bourgeois}},
  \bibinfo {author} {\bibfnamefont {J.}~\bibnamefont {Hruby}}, \bibinfo
  {author} {\bibfnamefont {P.}~\bibnamefont {Siyushev}}, \bibinfo {author}
  {\bibfnamefont {G.}~\bibnamefont {Wachter}}, \bibinfo {author} {\bibfnamefont
  {F.}~\bibnamefont {Aumayr}}, \bibinfo {author} {\bibfnamefont {P.~R.}\
  \bibnamefont {Hemmer}}, \bibinfo {author} {\bibfnamefont {A.}~\bibnamefont
  {Gali}}, \bibinfo {author} {\bibfnamefont {F.}~\bibnamefont {Jelezko}},
  \bibinfo {author} {\bibfnamefont {M.}~\bibnamefont {Trupke}}, \ and\ \bibinfo
  {author} {\bibfnamefont {M.}~\bibnamefont {Nesladek}},\ }\href {\doibase
  10.1103/PhysRevApplied.7.044032} {\bibfield  {journal} {\bibinfo  {journal}
  {Phys. Rev. Applied}\ }\textbf {\bibinfo {volume} {7}},\ \bibinfo {pages}
  {044032} (\bibinfo {year} {2017})}\BibitemShut {NoStop}%
\bibitem [{\citenamefont {Iakoubovskii}\ and\ \citenamefont
  {Adriaenssens}(2001)}]{Iakoubovskii01}%
  \BibitemOpen
  \bibfield  {author} {\bibinfo {author} {\bibfnamefont {K.}~\bibnamefont
  {Iakoubovskii}}\ and\ \bibinfo {author} {\bibfnamefont {G.~J.}\ \bibnamefont
  {Adriaenssens}},\ }\href@noop {} {\bibfield  {journal} {\bibinfo  {journal}
  {Journal of Physics: Condensed Matter}\ }\textbf {\bibinfo {volume} {13}},\
  \bibinfo {pages} {6015} (\bibinfo {year} {2001})}\BibitemShut {NoStop}%
\bibitem [{\citenamefont {Rondin}\ \emph {et~al.}(2010)\citenamefont {Rondin},
  \citenamefont {Dantelle}, \citenamefont {Slablab}, \citenamefont {Grosshans},
  \citenamefont {Treussart}, \citenamefont {Bergonzo}, \citenamefont
  {Perruchas}, \citenamefont {Gacoin}, \citenamefont {Chaigneau}, \citenamefont
  {Chang}, \citenamefont {Jacques},\ and\ \citenamefont {Roch}}]{ron10}%
  \BibitemOpen
  \bibfield  {author} {\bibinfo {author} {\bibfnamefont {L.}~\bibnamefont
  {Rondin}}, \bibinfo {author} {\bibfnamefont {G.}~\bibnamefont {Dantelle}},
  \bibinfo {author} {\bibfnamefont {A.}~\bibnamefont {Slablab}}, \bibinfo
  {author} {\bibfnamefont {F.}~\bibnamefont {Grosshans}}, \bibinfo {author}
  {\bibfnamefont {F.}~\bibnamefont {Treussart}}, \bibinfo {author}
  {\bibfnamefont {P.}~\bibnamefont {Bergonzo}}, \bibinfo {author}
  {\bibfnamefont {S.}~\bibnamefont {Perruchas}}, \bibinfo {author}
  {\bibfnamefont {T.}~\bibnamefont {Gacoin}}, \bibinfo {author} {\bibfnamefont
  {M.}~\bibnamefont {Chaigneau}}, \bibinfo {author} {\bibfnamefont {H.-C.}\
  \bibnamefont {Chang}}, \bibinfo {author} {\bibfnamefont {V.}~\bibnamefont
  {Jacques}}, \ and\ \bibinfo {author} {\bibfnamefont {J.-F.}\ \bibnamefont
  {Roch}},\ }\href {\doibase 10.1103/PhysRevB.82.115449} {\bibfield  {journal}
  {\bibinfo  {journal} {Phys. Rev. B}\ }\textbf {\bibinfo {volume} {82}},\
  \bibinfo {pages} {115449} (\bibinfo {year} {2010})}\BibitemShut {NoStop}%
\bibitem [{\citenamefont {Wee}\ \emph {et~al.}(2007)\citenamefont {Wee},
  \citenamefont {Tzeng}, \citenamefont {Han}, \citenamefont {Chang},
  \citenamefont {Fann}, \citenamefont {Hsu}, \citenamefont {Chen},\ and\
  \citenamefont {Yu}}]{wee2007}%
  \BibitemOpen
  \bibfield  {author} {\bibinfo {author} {\bibfnamefont {T.-L.}\ \bibnamefont
  {Wee}}, \bibinfo {author} {\bibfnamefont {Y.-K.}\ \bibnamefont {Tzeng}},
  \bibinfo {author} {\bibfnamefont {C.-C.}\ \bibnamefont {Han}}, \bibinfo
  {author} {\bibfnamefont {H.-C.}\ \bibnamefont {Chang}}, \bibinfo {author}
  {\bibfnamefont {W.}~\bibnamefont {Fann}}, \bibinfo {author} {\bibfnamefont
  {J.-H.}\ \bibnamefont {Hsu}}, \bibinfo {author} {\bibfnamefont {K.-M.}\
  \bibnamefont {Chen}}, \ and\ \bibinfo {author} {\bibfnamefont {Y.-C.}\
  \bibnamefont {Yu}},\ }\href {\doibase 10.1021/jp073938o} {\bibfield
  {journal} {\bibinfo  {journal} {The Journal of Physical Chemistry A}\
  }\textbf {\bibinfo {volume} {111}},\ \bibinfo {pages} {9379} (\bibinfo {year}
  {2007})},\ \bibinfo {note} {pMID: 17705460},\ \Eprint
  {http://arxiv.org/abs/http://dx.doi.org/10.1021/jp073938o}
  {http://dx.doi.org/10.1021/jp073938o} \BibitemShut {NoStop}%
\bibitem [{\citenamefont {Lawson}\ \emph {et~al.}(1998)\citenamefont {Lawson},
  \citenamefont {Fisher}, \citenamefont {Hunt},\ and\ \citenamefont
  {Newton}}]{lawson1998}%
  \BibitemOpen
  \bibfield  {author} {\bibinfo {author} {\bibfnamefont {S.~C.}\ \bibnamefont
  {Lawson}}, \bibinfo {author} {\bibfnamefont {D.}~\bibnamefont {Fisher}},
  \bibinfo {author} {\bibfnamefont {D.~C.}\ \bibnamefont {Hunt}}, \ and\
  \bibinfo {author} {\bibfnamefont {M.~E.}\ \bibnamefont {Newton}},\ }\href
  {http://stacks.iop.org/0953-8984/10/i=27/a=016} {\bibfield  {journal}
  {\bibinfo  {journal} {Journal of Physics: Condensed Matter}\ }\textbf
  {\bibinfo {volume} {10}},\ \bibinfo {pages} {6171} (\bibinfo {year}
  {1998})}\BibitemShut {NoStop}%
\bibitem [{\citenamefont {Collins}\ and\ \citenamefont
  {Spear}(1982)}]{collins82}%
  \BibitemOpen
  \bibfield  {author} {\bibinfo {author} {\bibfnamefont {A.~T.}\ \bibnamefont
  {Collins}}\ and\ \bibinfo {author} {\bibfnamefont {P.~M.}\ \bibnamefont
  {Spear}},\ }\href {http://stacks.iop.org/0022-3727/15/i=12/a=006} {\bibfield
  {journal} {\bibinfo  {journal} {Journal of Physics D: Applied Physics}\
  }\textbf {\bibinfo {volume} {15}},\ \bibinfo {pages} {L183} (\bibinfo {year}
  {1982})}\BibitemShut {NoStop}%
\bibitem [{\citenamefont {Lowther}(1995)}]{lowther1995}%
  \BibitemOpen
  \bibfield  {author} {\bibinfo {author} {\bibfnamefont {J.~E.}\ \bibnamefont
  {Lowther}},\ }\href {\doibase 10.1103/PhysRevB.51.91} {\bibfield  {journal}
  {\bibinfo  {journal} {Phys. Rev. B}\ }\textbf {\bibinfo {volume} {51}},\
  \bibinfo {pages} {91} (\bibinfo {year} {1995})}\BibitemShut {NoStop}%
\bibitem [{\citenamefont {Iakoubovskii}\ \emph {et~al.}(2000)\citenamefont
  {Iakoubovskii}, \citenamefont {Stesmans}, \citenamefont {Nouwen},\ and\
  \citenamefont {Adriaenssens}}]{iakoubovskii2000}%
  \BibitemOpen
  \bibfield  {author} {\bibinfo {author} {\bibfnamefont {K.}~\bibnamefont
  {Iakoubovskii}}, \bibinfo {author} {\bibfnamefont {A.}~\bibnamefont
  {Stesmans}}, \bibinfo {author} {\bibfnamefont {B.}~\bibnamefont {Nouwen}}, \
  and\ \bibinfo {author} {\bibfnamefont {G.~J.}\ \bibnamefont {Adriaenssens}},\
  }\href {\doibase 10.1103/PhysRevB.62.16587} {\bibfield  {journal} {\bibinfo
  {journal} {Phys. Rev. B}\ }\textbf {\bibinfo {volume} {62}},\ \bibinfo
  {pages} {16587} (\bibinfo {year} {2000})}\BibitemShut {NoStop}%
\bibitem [{\citenamefont {Goss}\ \emph {et~al.}(2004)\citenamefont {Goss},
  \citenamefont {Briddon}, \citenamefont {Jones},\ and\ \citenamefont
  {\"Oberg}}]{goss04}%
  \BibitemOpen
  \bibfield  {author} {\bibinfo {author} {\bibfnamefont {J.~P.}\ \bibnamefont
  {Goss}}, \bibinfo {author} {\bibfnamefont {P.~R.}\ \bibnamefont {Briddon}},
  \bibinfo {author} {\bibfnamefont {R.}~\bibnamefont {Jones}}, \ and\ \bibinfo
  {author} {\bibfnamefont {S.}~\bibnamefont {\"Oberg}},\ }\href
  {http://stacks.iop.org/0953-8984/16/i=25/a=014} {\bibfield  {journal}
  {\bibinfo  {journal} {Journal of Physics: Condensed Matter}\ }\textbf
  {\bibinfo {volume} {16}},\ \bibinfo {pages} {4567} (\bibinfo {year}
  {2004})}\BibitemShut {NoStop}%
\bibitem [{\citenamefont {Larico}\ \emph {et~al.}(2009)\citenamefont {Larico},
  \citenamefont {Justo}, \citenamefont {Machado},\ and\ \citenamefont
  {Assali}}]{larico09}%
  \BibitemOpen
  \bibfield  {author} {\bibinfo {author} {\bibfnamefont {R.}~\bibnamefont
  {Larico}}, \bibinfo {author} {\bibfnamefont {J.~F.}\ \bibnamefont {Justo}},
  \bibinfo {author} {\bibfnamefont {W.~V.~M.}\ \bibnamefont {Machado}}, \ and\
  \bibinfo {author} {\bibfnamefont {L.~V.~C.}\ \bibnamefont {Assali}},\ }\href
  {\doibase 10.1103/PhysRevB.79.115202} {\bibfield  {journal} {\bibinfo
  {journal} {Phys. Rev. B}\ }\textbf {\bibinfo {volume} {79}},\ \bibinfo
  {pages} {115202} (\bibinfo {year} {2009})}\BibitemShut {NoStop}%
\bibitem [{\citenamefont {Thiering}\ \emph {et~al.}(2014)\citenamefont
  {Thiering}, \citenamefont {Londero},\ and\ \citenamefont {Gali}}]{thiering}%
  \BibitemOpen
  \bibfield  {author} {\bibinfo {author} {\bibfnamefont {G.}~\bibnamefont
  {Thiering}}, \bibinfo {author} {\bibfnamefont {E.}~\bibnamefont {Londero}}, \
  and\ \bibinfo {author} {\bibfnamefont {A.}~\bibnamefont {Gali}},\ }\href
  {\doibase 10.1039/C4NR03112A} {\bibfield  {journal} {\bibinfo  {journal}
  {Nanoscale}\ }\textbf {\bibinfo {volume} {6}},\ \bibinfo {pages} {12018}
  (\bibinfo {year} {2014})}\BibitemShut {NoStop}%
\bibitem [{\citenamefont {Yelisseyev}\ and\ \citenamefont
  {Nadolinny}(1995)}]{Yelisseyev95}%
  \BibitemOpen
  \bibfield  {author} {\bibinfo {author} {\bibfnamefont {A.~P.}\ \bibnamefont
  {Yelisseyev}}\ and\ \bibinfo {author} {\bibfnamefont {V.~A.}\ \bibnamefont
  {Nadolinny}},\ }\href {\doibase 10.1016/0925-9635(94)00240-1} {\bibfield
  {journal} {\bibinfo  {journal} {Diamond and Related Materials}\ }\textbf
  {\bibinfo {volume} {4}},\ \bibinfo {pages} {177} (\bibinfo {year}
  {1995})}\BibitemShut {NoStop}%
\bibitem [{\citenamefont {Gali}\ \emph {et~al.}(2009)\citenamefont {Gali},
  \citenamefont {Janz\'en}, \citenamefont {De\'ak}, \citenamefont {Kresse},\
  and\ \citenamefont {Kaxiras}}]{gali09}%
  \BibitemOpen
  \bibfield  {author} {\bibinfo {author} {\bibfnamefont {A.}~\bibnamefont
  {Gali}}, \bibinfo {author} {\bibfnamefont {E.}~\bibnamefont {Janz\'en}},
  \bibinfo {author} {\bibfnamefont {P.}~\bibnamefont {De\'ak}}, \bibinfo
  {author} {\bibfnamefont {G.}~\bibnamefont {Kresse}}, \ and\ \bibinfo {author}
  {\bibfnamefont {E.}~\bibnamefont {Kaxiras}},\ }\href {\doibase
  10.1103/PhysRevLett.103.186404} {\bibfield  {journal} {\bibinfo  {journal}
  {Phys. Rev. Lett.}\ }\textbf {\bibinfo {volume} {103}},\ \bibinfo {pages}
  {186404} (\bibinfo {year} {2009})}\BibitemShut {NoStop}%
\bibitem [{\citenamefont {De\'ak}\ \emph {et~al.}(2010)\citenamefont {De\'ak},
  \citenamefont {Aradi}, \citenamefont {Frauenheim}, \citenamefont {Janz\'en},\
  and\ \citenamefont {Gali}}]{deak}%
  \BibitemOpen
  \bibfield  {author} {\bibinfo {author} {\bibfnamefont {P.}~\bibnamefont
  {De\'ak}}, \bibinfo {author} {\bibfnamefont {B.}~\bibnamefont {Aradi}},
  \bibinfo {author} {\bibfnamefont {T.}~\bibnamefont {Frauenheim}}, \bibinfo
  {author} {\bibfnamefont {E.}~\bibnamefont {Janz\'en}}, \ and\ \bibinfo
  {author} {\bibfnamefont {A.}~\bibnamefont {Gali}},\ }\href {\doibase
  10.1103/PhysRevB.81.153203} {\bibfield  {journal} {\bibinfo  {journal} {Phys.
  Rev. B}\ }\textbf {\bibinfo {volume} {81}},\ \bibinfo {pages} {153203}
  (\bibinfo {year} {2010})}\BibitemShut {NoStop}%
\bibitem [{\citenamefont {Heyd}\ \emph {et~al.}(2003)\citenamefont {Heyd},
  \citenamefont {Scuseria},\ and\ \citenamefont {Ernzerhof}}]{heyd}%
  \BibitemOpen
  \bibfield  {author} {\bibinfo {author} {\bibfnamefont {J.}~\bibnamefont
  {Heyd}}, \bibinfo {author} {\bibfnamefont {G.~E.}\ \bibnamefont {Scuseria}},
  \ and\ \bibinfo {author} {\bibfnamefont {M.}~\bibnamefont {Ernzerhof}},\
  }\href {\doibase http://dx.doi.org/10.1063/1.1564060} {\bibfield  {journal}
  {\bibinfo  {journal} {The Journal of Chemical Physics}\ }\textbf {\bibinfo
  {volume} {118}},\ \bibinfo {eid} {8207} (\bibinfo {year} {2003})}\BibitemShut
  {NoStop}%
\bibitem [{\citenamefont {Krukau}\ \emph {et~al.}(2006)\citenamefont {Krukau},
  \citenamefont {Vydrov}, \citenamefont {Izmaylov},\ and\ \citenamefont
  {Scuseria}}]{krukau}%
  \BibitemOpen
  \bibfield  {author} {\bibinfo {author} {\bibfnamefont {A.~V.}\ \bibnamefont
  {Krukau}}, \bibinfo {author} {\bibfnamefont {O.~A.}\ \bibnamefont {Vydrov}},
  \bibinfo {author} {\bibfnamefont {A.~F.}\ \bibnamefont {Izmaylov}}, \ and\
  \bibinfo {author} {\bibfnamefont {G.~E.}\ \bibnamefont {Scuseria}},\ }\href
  {\doibase http://dx.doi.org/10.1063/1.2404663} {\bibfield  {journal}
  {\bibinfo  {journal} {The Journal of Chemical Physics}\ }\textbf {\bibinfo
  {volume} {125}},\ \bibinfo {eid} {224106} (\bibinfo {year}
  {2006})}\BibitemShut {NoStop}%
\bibitem [{\citenamefont {Perdew}\ \emph {et~al.}(1996)\citenamefont {Perdew},
  \citenamefont {Burke},\ and\ \citenamefont {Ernzerhof}}]{perdew96}%
  \BibitemOpen
  \bibfield  {author} {\bibinfo {author} {\bibfnamefont {J.~P.}\ \bibnamefont
  {Perdew}}, \bibinfo {author} {\bibfnamefont {K.}~\bibnamefont {Burke}}, \
  and\ \bibinfo {author} {\bibfnamefont {M.}~\bibnamefont {Ernzerhof}},\ }\href
  {\doibase 10.1103/PhysRevLett.77.3865} {\bibfield  {journal} {\bibinfo
  {journal} {Phys. Rev. Lett.}\ }\textbf {\bibinfo {volume} {77}},\ \bibinfo
  {pages} {3865} (\bibinfo {year} {1996})}\BibitemShut {NoStop}%
\bibitem [{\citenamefont {Kresse}\ and\ \citenamefont {Hafner}(1993)}]{VASP1}%
  \BibitemOpen
  \bibfield  {author} {\bibinfo {author} {\bibfnamefont {G.}~\bibnamefont
  {Kresse}}\ and\ \bibinfo {author} {\bibfnamefont {J.}~\bibnamefont
  {Hafner}},\ }\href {\doibase 10.1103/PhysRevB.47.558} {\bibfield  {journal}
  {\bibinfo  {journal} {Phys. Rev. B}\ }\textbf {\bibinfo {volume} {47}},\
  \bibinfo {pages} {558} (\bibinfo {year} {1993})}\BibitemShut {NoStop}%
\bibitem [{\citenamefont {Kresse}\ and\ \citenamefont
  {Furthm\"uller}(1996)}]{VASP2}%
  \BibitemOpen
  \bibfield  {author} {\bibinfo {author} {\bibfnamefont {G.}~\bibnamefont
  {Kresse}}\ and\ \bibinfo {author} {\bibfnamefont {J.}~\bibnamefont
  {Furthm\"uller}},\ }\href {\doibase 10.1103/PhysRevB.54.11169} {\bibfield
  {journal} {\bibinfo  {journal} {Phys. Rev. B}\ }\textbf {\bibinfo {volume}
  {54}},\ \bibinfo {pages} {11169} (\bibinfo {year} {1996})}\BibitemShut
  {NoStop}%
\bibitem [{\citenamefont {Bl\"ochl}(1994)}]{paw}%
  \BibitemOpen
  \bibfield  {author} {\bibinfo {author} {\bibfnamefont {P.~E.}\ \bibnamefont
  {Bl\"ochl}},\ }\href {\doibase 10.1103/PhysRevB.50.17953} {\bibfield
  {journal} {\bibinfo  {journal} {Phys. Rev. B}\ }\textbf {\bibinfo {volume}
  {50}},\ \bibinfo {pages} {17953} (\bibinfo {year} {1994})}\BibitemShut
  {NoStop}%
\bibitem [{\citenamefont {Freysoldt}\ \emph {et~al.}(2009)\citenamefont
  {Freysoldt}, \citenamefont {Neugebauer},\ and\ \citenamefont {Van~de
  Walle}}]{freysoldt1}%
  \BibitemOpen
  \bibfield  {author} {\bibinfo {author} {\bibfnamefont {C.}~\bibnamefont
  {Freysoldt}}, \bibinfo {author} {\bibfnamefont {J.}~\bibnamefont
  {Neugebauer}}, \ and\ \bibinfo {author} {\bibfnamefont {C.~G.}\ \bibnamefont
  {Van~de Walle}},\ }\href {\doibase 10.1103/PhysRevLett.102.016402} {\bibfield
   {journal} {\bibinfo  {journal} {Phys. Rev. Lett.}\ }\textbf {\bibinfo
  {volume} {102}},\ \bibinfo {pages} {016402} (\bibinfo {year}
  {2009})}\BibitemShut {NoStop}%
\bibitem [{\citenamefont {De\'ak}\ \emph {et~al.}(2014)\citenamefont {De\'ak},
  \citenamefont {Aradi}, \citenamefont {Kaviani}, \citenamefont {Frauenheim},\
  and\ \citenamefont {Gali}}]{deak2014}%
  \BibitemOpen
  \bibfield  {author} {\bibinfo {author} {\bibfnamefont {P.}~\bibnamefont
  {De\'ak}}, \bibinfo {author} {\bibfnamefont {B.}~\bibnamefont {Aradi}},
  \bibinfo {author} {\bibfnamefont {M.}~\bibnamefont {Kaviani}}, \bibinfo
  {author} {\bibfnamefont {T.}~\bibnamefont {Frauenheim}}, \ and\ \bibinfo
  {author} {\bibfnamefont {A.}~\bibnamefont {Gali}},\ }\href {\doibase
  10.1103/PhysRevB.89.075203} {\bibfield  {journal} {\bibinfo  {journal} {Phys.
  Rev. B}\ }\textbf {\bibinfo {volume} {89}},\ \bibinfo {pages} {075203}
  (\bibinfo {year} {2014})}\BibitemShut {NoStop}%
\bibitem [{SM()}]{SM}%
  \BibitemOpen
  \href@noop {} {}\bibinfo {howpublished} {See Supplemental Material at
  http://link.aps.org/supplemental/ 10.1103/PhysRevB.xx.xxxxxx for details on
  the ionization energies of the nitrogen related defects in
  diamond.}\BibitemShut {Stop}%
\bibitem [{\citenamefont {Nesladek}\ \emph {et~al.}(1998)\citenamefont
  {Nesladek}, \citenamefont {Stals}, \citenamefont {Stesmans}, \citenamefont
  {Iakoubovskij}, \citenamefont {Adriaenssens}, \citenamefont {Rosa},\ and\
  \citenamefont {Van\v{e}\v{c}ek}}]{nesladek98}%
  \BibitemOpen
  \bibfield  {author} {\bibinfo {author} {\bibfnamefont {M.}~\bibnamefont
  {Nesladek}}, \bibinfo {author} {\bibfnamefont {L.~M.}\ \bibnamefont {Stals}},
  \bibinfo {author} {\bibfnamefont {A.}~\bibnamefont {Stesmans}}, \bibinfo
  {author} {\bibfnamefont {K.}~\bibnamefont {Iakoubovskij}}, \bibinfo {author}
  {\bibfnamefont {G.~J.}\ \bibnamefont {Adriaenssens}}, \bibinfo {author}
  {\bibfnamefont {J.}~\bibnamefont {Rosa}}, \ and\ \bibinfo {author}
  {\bibfnamefont {M.}~\bibnamefont {Van\v{e}\v{c}ek}},\ }\href {\doibase
  10.1063/1.121632} {\bibfield  {journal} {\bibinfo  {journal} {Applied Physics
  Letters}\ }\textbf {\bibinfo {volume} {72}},\ \bibinfo {pages} {3306}
  (\bibinfo {year} {1998})}\BibitemShut {NoStop}%
\bibitem [{\citenamefont {Rosa}\ \emph {et~al.}(1999)\citenamefont {Rosa},
  \citenamefont {Van\v{e}\v{c}ek}, \citenamefont {Nesladek},\ and\
  \citenamefont {Stals}}]{Rosa99}%
  \BibitemOpen
  \bibfield  {author} {\bibinfo {author} {\bibfnamefont {J.}~\bibnamefont
  {Rosa}}, \bibinfo {author} {\bibfnamefont {M.}~\bibnamefont
  {Van\v{e}\v{c}ek}}, \bibinfo {author} {\bibfnamefont {M.}~\bibnamefont
  {Nesladek}}, \ and\ \bibinfo {author} {\bibfnamefont {L.}~\bibnamefont
  {Stals}},\ }\href {\doibase http://dx.doi.org/10.1016/S0925-9635(98)00354-9}
  {\bibfield  {journal} {\bibinfo  {journal} {Diamond and Related Materials}\
  }\textbf {\bibinfo {volume} {8}},\ \bibinfo {pages} {721 } (\bibinfo {year}
  {1999})}\BibitemShut {NoStop}%
\bibitem [{\citenamefont {Inkson}(1981)}]{inkson81}%
  \BibitemOpen
  \bibfield  {author} {\bibinfo {author} {\bibfnamefont {J.~C.}\ \bibnamefont
  {Inkson}},\ }\href {http://stacks.iop.org/0022-3719/14/i=7/a=012} {\bibfield
  {journal} {\bibinfo  {journal} {Journal of Physics C: Solid State Physics}\
  }\textbf {\bibinfo {volume} {14}},\ \bibinfo {pages} {1093} (\bibinfo {year}
  {1981})}\BibitemShut {NoStop}%
\bibitem [{\citenamefont {Lawson}\ \emph {et~al.}(1993)\citenamefont {Lawson},
  \citenamefont {Kanda},\ and\ \citenamefont {Sekita}}]{lawson1993}%
  \BibitemOpen
  \bibfield  {author} {\bibinfo {author} {\bibfnamefont {S.~C.}\ \bibnamefont
  {Lawson}}, \bibinfo {author} {\bibfnamefont {H.}~\bibnamefont {Kanda}}, \
  and\ \bibinfo {author} {\bibfnamefont {M.}~\bibnamefont {Sekita}},\ }\href
  {\doibase 10.1080/13642819308215280} {\bibfield  {journal} {\bibinfo
  {journal} {Philosophical Magazine Part B}\ }\textbf {\bibinfo {volume}
  {68}},\ \bibinfo {pages} {39} (\bibinfo {year} {1993})}\BibitemShut {NoStop}%
\bibitem [{\citenamefont {Nadolinny}\ \emph {et~al.}(2002)\citenamefont
  {Nadolinny}, \citenamefont {Baker}, \citenamefont {Newton},\ and\
  \citenamefont {Kanda}}]{nadolinny2002}%
  \BibitemOpen
  \bibfield  {author} {\bibinfo {author} {\bibfnamefont {V.~A.}\ \bibnamefont
  {Nadolinny}}, \bibinfo {author} {\bibfnamefont {J.~M.}\ \bibnamefont
  {Baker}}, \bibinfo {author} {\bibfnamefont {M.~E.}\ \bibnamefont {Newton}}, \
  and\ \bibinfo {author} {\bibfnamefont {H.}~\bibnamefont {Kanda}},\ }\href
  {\doibase 10.1016/S0925-9635(01)00620-3} {\bibfield  {journal} {\bibinfo
  {journal} {Diamond and Related Materials}\ }\bibinfo {series} {12th
  {European} {Conference} on {Diamond}, {Diamond}- {Like} {Materials}, {Carbon}
  {Nanotubes}, {Nitrides} \& {Silicon} {Carbide}},\ \textbf {\bibinfo {volume}
  {11}},\ \bibinfo {pages} {627} (\bibinfo {year} {2002})}\BibitemShut
  {NoStop}%
\bibitem [{\citenamefont {Iakoubovskii}(2004)}]{Iakoubovskii04}%
  \BibitemOpen
  \bibfield  {author} {\bibinfo {author} {\bibfnamefont {K.}~\bibnamefont
  {Iakoubovskii}},\ }\href {\doibase 10.1103/PhysRevB.70.205211} {\bibfield
  {journal} {\bibinfo  {journal} {Phys. Rev. B}\ }\textbf {\bibinfo {volume}
  {70}},\ \bibinfo {pages} {205211} (\bibinfo {year} {2004})}\BibitemShut
  {NoStop}%
\bibitem [{\citenamefont {Iv\'ady}\ \emph {et~al.}(2014)\citenamefont
  {Iv\'ady}, \citenamefont {Simon}, \citenamefont {Maze}, \citenamefont
  {Abrikosov},\ and\ \citenamefont {Gali}}]{ivady2014}%
  \BibitemOpen
  \bibfield  {author} {\bibinfo {author} {\bibfnamefont {V.}~\bibnamefont
  {Iv\'ady}}, \bibinfo {author} {\bibfnamefont {T.}~\bibnamefont {Simon}},
  \bibinfo {author} {\bibfnamefont {J.~R.}\ \bibnamefont {Maze}}, \bibinfo
  {author} {\bibfnamefont {I.~A.}\ \bibnamefont {Abrikosov}}, \ and\ \bibinfo
  {author} {\bibfnamefont {A.}~\bibnamefont {Gali}},\ }\href {\doibase
  10.1103/PhysRevB.90.235205} {\bibfield  {journal} {\bibinfo  {journal} {Phys.
  Rev. B}\ }\textbf {\bibinfo {volume} {90}},\ \bibinfo {pages} {235205}
  (\bibinfo {year} {2014})}\BibitemShut {NoStop}%
\bibitem [{\citenamefont {Sz\'asz}\ \emph {et~al.}(2013)\citenamefont
  {Sz\'asz}, \citenamefont {Hornos}, \citenamefont {Marsman},\ and\
  \citenamefont {Gali}}]{szasz2013}%
  \BibitemOpen
  \bibfield  {author} {\bibinfo {author} {\bibfnamefont {K.}~\bibnamefont
  {Sz\'asz}}, \bibinfo {author} {\bibfnamefont {T.}~\bibnamefont {Hornos}},
  \bibinfo {author} {\bibfnamefont {M.}~\bibnamefont {Marsman}}, \ and\
  \bibinfo {author} {\bibfnamefont {A.}~\bibnamefont {Gali}},\ }\href {\doibase
  10.1103/PhysRevB.88.075202} {\bibfield  {journal} {\bibinfo  {journal} {Phys.
  Rev. B}\ }\textbf {\bibinfo {volume} {88}},\ \bibinfo {pages} {075202}
  (\bibinfo {year} {2013})}\BibitemShut {NoStop}%
\bibitem [{\citenamefont {Edmonds}\ \emph {et~al.}(2008)\citenamefont
  {Edmonds}, \citenamefont {Newton}, \citenamefont {Martineau}, \citenamefont
  {Twitchen},\ and\ \citenamefont {Williams}}]{edmonds08}%
  \BibitemOpen
  \bibfield  {author} {\bibinfo {author} {\bibfnamefont {A.~M.}\ \bibnamefont
  {Edmonds}}, \bibinfo {author} {\bibfnamefont {M.~E.}\ \bibnamefont {Newton}},
  \bibinfo {author} {\bibfnamefont {P.~M.}\ \bibnamefont {Martineau}}, \bibinfo
  {author} {\bibfnamefont {D.~J.}\ \bibnamefont {Twitchen}}, \ and\ \bibinfo
  {author} {\bibfnamefont {S.~D.}\ \bibnamefont {Williams}},\ }\href {\doibase
  10.1103/PhysRevB.77.245205} {\bibfield  {journal} {\bibinfo  {journal} {Phys.
  Rev. B}\ }\textbf {\bibinfo {volume} {77}},\ \bibinfo {pages} {245205}
  (\bibinfo {year} {2008})}\BibitemShut {NoStop}%
\end{thebibliography}

%

\end{document}